\documentclass[aip,pop,preprint,linenumber,superscriptaddress,amsmath,amsfonts]{revtex4-1}
\usepackage{color,graphicx} 
\usepackage{multirow}
\usepackage{ifthen}
\usepackage{ulem} 
\usepackage{bm}

\usepackage[colorlinks,linkcolor=blue,citecolor=blue,bookmarks,pdfstartview=FitH]{hyperref}

\newcommand{\myfigdir}[1]{}
\newcommand{\revcolor}[1]{{\color{black}#1}}
\newcommand{\revcolora}[1]{{\color{black}#1}}
\newcommand{\revcolorb}[1]{{\color{black}#1}}

\begin{document}
	\title{The development of an implicit full $f$  method for electromagnetic particle simulations of Alfv\'en waves and energetic particle physics}
	\author{Z. X. Lu}
	\email{zhixin.lu@ipp.mpg.de}
	\homepage{http://www2.ipp.mpg.de/~luzhixin/}
	\author{G. Meng}
	\author{M. Hoelzl}
	\author{Ph. Lauber}
	\affiliation{Max-Planck-Institut f\"ur Plasmaphysik,  85748 Garching, Germany}
	
	
	\begin{abstract}
		In this work, an implicit scheme for particle-in-cell/Fourier electromagnetic simulations is developed and applied to studies of Alfv\'en waves in one dimension and \revcolor{three dimensional} tokamak plasmas. An analytical treatment is introduced to achieve efficient convergence of the iterative solution of the implicit field-particle system. \revcolorb{First}, its application to the one-dimensional uniform plasma demonstrates its applicability in a broad range of $\beta/m_e$ values. 
		\revcolorb{Second, toroidicity induced Alfv\'en eigenmodes (TAE) are simulated in a three dimensional axisymmetric tokamak plasma,} using the widely studied case defined by the International Tokamak Physics Activity (ITPA) Energetic Particle (EP) Topical Group. The real frequency and the growth (or damping) rate of the TAE with (or without) EPs agree with previous results reasonably well. The full $f$  electromagnetic particle scheme established in this work provides a possible natural choice for EP transport studies where large profile variation \revcolor{and arbitrary particle distribution functions need to be treated} in kinetic simulations. 
	\end{abstract}

	\maketitle
	
	\section{Introduction}
	The gyrokinetic particle-in-cell (PIC) simulation serves as a 
	\revcolorb{tool close to first principles} for the studies of tokamak plasmas \cite{lee1983gyrokinetic}, and has revealed the importance of the zonal flow \cite{lin1998turbulent}, the kinetic \revcolora{properties} of energetic particles \cite{wang2013radial} and the edge transport features\cite{chang2017fast}. While most gyrokinetic particle codes are based on explicit time stepping and the $\delta f$ method, \revcolor{where $\delta f$ is the perturbed distribution function} \cite{parker1993fully}, the implicit PIC method in slab geometry has been reported featured with good properties such as energy and momentum conservation and the capability of allowing large time steps \cite{chen2011energy}. 
	\revcolorb{In addition, the full $f$ method does not rely on the separation of the equilibrium and the perturbation, and thus provides a natural way to handle substantial changes of the profiles in the course of a simulation \cite{heikkinen2008full},  while for the $\delta f$ scheme, the advantages in noise control are lost to some extent in such a scenario  and the positive-definiteness of the distribution function needs to be ensured as $\delta f$ and $f_0$ approach similar orders of magnitude, where $f_0=f-\delta f$. A full $f$ scheme can be easily applied to arbitrary distribution functions, without calculating the phase space derivatives of the equilibrium distribution function $f_0$ as required in the $\delta f$ method. }
	In the study of MHD/fluid problems, the mixed explicit-implicit scheme has been developed \cite{gunter2009mixed}, which shed some light on the development of gyrokinetic or hybrid particle-fluid method (kinetic MHD). 
	One crucial issue in both fluid problem and kinetic problem is to treat the parallel dynamics accurately, considering the distinct features in parallel and perpendicular direction such as the large parallel to perpendicular transport coefficients ratio and, when kinetic particles are included, the fast response of electrons in the parallel direction \cite{mishchenko2019pullback}. 
	While the pullback scheme is developed successfully for the electromagnetic simulation, 
	\revcolora{it is shown that a similar linear numerical dispersion relation can be obtained using the implicit scheme based on a simplified model in slab geometry, without analyses  of the particle noise levels and computational costs in the derivation \cite{kleiber2016explicit}, which indicates that with the same time step size $\Delta t$, similar frequencies and damping rates can be obtained by either using the pullback scheme or the implicit scheme in the linear limit.}
	\revcolora{Generally, the implicit scheme is known for its capability of allowing large time steps \cite{cohen1989performance,chen2011energy}. Moreover, with a specific discrete formulation, the implicit scheme can ensure} good conservation properties \cite{chen2011energy}.  
	\revcolora{As in the widely used electromagnetic gyrokinetic model, the electrostatic and electromagnetic potentials $\delta\phi$ and $\delta {\bm A}$ are chosen as variables \cite{brizard2007foundations,chen2001gyrokinetic,sturdevant2019fully}. In the ``symplectic ($v_\parallel$)'' formula, the parallel velocity ($v_\parallel$) of the particles' guiding center is adopted and numerical challenges arise due to the $\partial\delta{A}_\parallel/\partial t$ term in the $d v_\parallel/dt$ equation. In the ``Hamiltionian ($p_\parallel$)'' formula, $\partial\delta{A}_\parallel/\partial t$ is eliminated but the ``cancellation'' problem appears \cite{chen2001gyrokinetic,hatzky2007electromagnetic}. The implicit scheme provides a natural treatment of the $\partial\delta{A}_\parallel/\partial t$ term in the ``symplectic ($v_\parallel$)'' formula.} 
	The applications of the implicit scheme in the simulation of the \revcolora{electrostatic} toroidal ion temperature gradient instability \revcolor{have} been reported  \cite{sturdevant2017low} and a fully implicit scheme is studied recently in the particle simulation code XGC \cite{sturdevant2019fully}. 
	Nevertheless, the development and the application of the implicit full $f$  scheme on the study of Alfv\'en modes and energetic particle (EP) physics in tokamak plasmas have not been reported. 
	
	In this work, an implicit scheme for particle simulations is developed and implemented in TRIMEG-GKX. Instead of solving the implicit field-particle system numerically \cite{sturdevant2019fully}, we developed the analytical expansion for solving the implicit solution in order to generate the linear system, whose solution converges to that of the nonlinear system. This scheme is applied to the study of the Shear Alfv\'en Wave (SAW) \revcolor{in one dimension} and the Toroidicity induced Alfv\'en Eigenmode (TAE) excited by the \revcolora{energetic} particles 
	\revcolorb{in three dimensional axisymmetric tokamak plasmas}. 
	This work aims at providing
	\begin{enumerate}
		\item a demonstration of the applicability of the implicit method for the study of the SAW in tokamak plasmas;
		\item a mixed implicit-explicit scheme for particle simulations, with analytical simplifications, as a practical way to upgrade the TRIMEG code \cite{lu2019development}, meanwhile also as a potential solution for JOREK and other existing codes \cite{briguglio1995hybrid,lanti2019orb5,chang2017fast,huysmans2007mhd}, for dealing with full $f$  electromagnetic simulations;
		\item a full $f$  numerical tool for the study of Alfv\'en waves and energetic particle physics \cite{chen2016physics} that can deal with strong profile changes \revcolor{and arbitrary particle distribution functions  in a natural way, which}  is complementing existing codes\cite{lauber2007ligka,lanti2019orb5,wang2013radial}.
	\end{enumerate}
	This paper is organized as follows. In Section \ref{sec:model}, the model for the electromagnetic particle simulation is introduced. In Section \ref{sec:numerical}, the implicit scheme with analytical treatment is derived. In Section \ref{sec:result}, the simulation results of SAW in slab geometry and the TAE in tokamak plasmas are shown. In Section \ref{sec:summery}, we  provide summary and outlook. 
	
	\section{Electromagnetic model}
	\label{sec:model}
	In this section, the electromagnetic model is presented. In order to understand the performance and the applicability of this  implicit scheme with analytical treatment, we introduce the equations for the electromagnetic simulations in general geometry and its reduction to one dimension. Furthermore, the normalization and the mixed particle-in-cell-particle-in-Fourier (PIC-PIF) scheme are introduced. 
	
	For the tokamak geometry, the coordinates $(r,\phi,\theta)$ are adopted and the magnetic field is represented as ${\bm B}=\nabla\psi\times\nabla\phi+F\nabla\phi$, \revcolora{ where $r,\phi,\theta$ are the radial, poloidal and toroidal coordinates, $\psi$ is the poloidal magnetic flux function and $F$ is the poloidal current function}. An ad hoc equilibrium has been adopted, featured with concentric circular magnetic flux surfaces and constant $F$. Neverthless, the scheme in this work is general, and it can be readily extended to arbitrary tokamak geometry. 
	
	\subsection{Gyrokinetic Vlasov-Poisson equations and the parallel electron dynamics}
	The gyrokinetic Poisson-Amp\'ere system is described as follows,
	\begin{eqnarray}
	-\nabla_\perp\cdot\sum_s\frac{n_{0s}e_s}{\omega_cB}\nabla_\perp\delta\phi & = & \sum_s e_s\delta n_s \;\;, \\
	-\nabla^2_\perp\delta A_\parallel & = & \mu_0 \sum_s\delta j_{\parallel,s} \;\;,
	\end{eqnarray}
	where $\omega_c=e_s B/m_s$, the subscript `$s$' and `$\parallel$' indicate the species `$s$' and the component parallel to the equilibrium magnetic field respectively, and $\mu_0$ is the vacuum permeability. 
	
	The guiding center's equations of motion are as follows,
	\begin{eqnarray}
	\label{eq:dRdt}
	\frac{d}{dt}{{\bm{R}}} &=& {\bm v}_{\parallel}+{\bm v}_d+\delta{\bm v}\;\;,\\
	\label{eq:dvpardt}
	\frac{d}{dt}v_\parallel &=& \dot{v}_{\parallel0} + \delta\dot{v}_{\parallel}\;\;, \\
	\label{eq:vd}
	{\bm v}_d &=& \frac{m_s}{e_sB^2}\left(v^2_\parallel+\mu B\right){\bm b}\times\nabla B \;\;, \\
	\label{eq:deltavEB}
	\delta{\bm v} &=& \frac{\bm b}{B}\times\nabla(\delta\phi-v_\parallel\delta A_\parallel)\;\;, \\
	\dot{v}_{\parallel0} &=& -\mu\partial_{\parallel}B\;\;,\\
	\label{eq:ddeltavpardt}
	\delta\dot{v}_{\parallel} &=& -\frac{e_s}{m_s}\left(\partial_\parallel\delta\phi+\partial_t\delta A_\parallel\right) \;\;,
	\end{eqnarray}
	where the magnetic moment $\mu=v_\perp^2/(2B)$, $v_\perp$ is the perpendicular velocity, ${\bm b}={\bm B}/B$. In order to minimize the technical complexity of the code implementation and to focus on the implicit scheme and the physics, we have ignored the finite Lamor radius effect and the higher order terms $\sim\rho_{s}/L_B$, compared with the more comprehensive gyrokinetic model \cite{lin1998turbulent,chang2017fast,mishchenko2019pullback}, where $\rho_{s}=v_\perp/\omega_c$ is the Lamor radius of the particle species `$s$', and $L_B$ is the characteristic length of the equilibrium magnetic field. In spite of the simplification, it can be shown that the energy $E=v^2/2$ and the canonical toroidal angular momentum $P_\phi=e_s\psi+mv_\parallel F/B$ are conserved for the guiding center motion in equilibrium, i.e.,
	\begin{eqnarray}
	\frac{d}{dt}E_0=0\;\;,\;\; \frac{d}{dt}P_{\phi0}=0\;\;,
	\end{eqnarray}
	where the subscript `0' indicates the variables in equilibrium magnetic field. 
	
	For the one dimensional (1D) case, we consider the guiding center motion in uniform magnetic field (${\bm v}_d=0$, $\dot{v}_{\parallel0}=0$). In addition, we assume uniform density and temperature 
	\revcolorb{in all directions, and thus ${\bm b}\times\nabla(\delta\phi-v\delta{A}_\parallel)\cdot\nabla f_0$ vanishes in the linear dispersion relation, yielding $(\partial_t+v_\parallel\partial_{\parallel}+\delta\dot{v}_{\parallel}\partial/\partial v_\parallel)\delta f
	=-\delta{\bm v}\cdot\nabla f_0-\delta\dot{v}_\parallel\partial f_0/\partial v_{\parallel}
	=-\delta\dot{v}_\parallel\partial f_0/\partial v_{\parallel}$, 
	}where $f=f_0+\delta f$, $f_0$ and $\delta f$ are the equilibrium and the perturbed distribution functions respectively. Equations \ref{eq:dRdt} -- \ref{eq:ddeltavpardt} for the guiding center are reduced to 
	\begin{eqnarray}
	\frac{dl}{dt} &=& v_\parallel\;\;,\\
	\frac{dv_\parallel}{dt} &=& -\frac{e_s}{m_s} \left( \partial_\parallel\delta\phi+\partial_t\delta A_\parallel \right) \;\;,
	\end{eqnarray}
	where $l$ is the coordinate along the magnetic field.
	This 1D model is a good test case for the implicit scheme, since the most numerically challenging term $\partial_t\delta{A}_\parallel$ is retained. The numerical scheme that applies to this 1D model can be readily extended for the tokamak geometry, for treating the $\partial_t\delta A_\parallel$ term. 
	
	\subsection{Normalization}
	The normalization units of the length and the time are \revcolorb{$R_N=1\;\text{m}$,} 
	$t_N=R_N/v_{N}$, where $v_{N}=\sqrt{2T_N/m_N}$, $m_N$ is the proton mass, $T_N$ is the reference temperature, chosen to be the on-axis ion temperature in this work. 
	\revcolora{Meter is chosen as the length unit, as is adopted in the field solver and particle pusher of the gyrokinetic simulation code GTS \cite{wang2006gyro}. The purpose of this choice is to be consistent with the EFIT equilibrium interface and the mesh generator in TRIMEG \cite{lu2019development} where meter is also used for the description of the geometry. In addition, while the Larmor radius is a natural choice for microturbulence studies, macroscopic instabilies can be excited by EPs and thus a macroscopic length (1 meter) is also a reasonable length unit.}
	Other variables are normalized using $v_N$, $t_N \ldots$, i.e., $v_\parallel=\bar{v}_\parallel v_N$, $R=\bar{R}R_N$. In the following, for the sake of simplicity, the bar  $\bar{\ldots}$ is omitted when no confusion is introduced. 
	
	The normalized field equations are as follows,
	\begin{eqnarray}
	\label{eq:poisson_torus}
	-\nabla_\perp\cdot g\nabla_\perp\delta \phi & = & C_P  \delta N \;\;, \\
	\label{eq:ampere_torus}
	-\nabla^2_\perp\delta  {A}_\parallel & = & C_A \delta J_\parallel \;\;, \\
	\label{eq:definegNJ}
	g=\sum_sM_s\frac{n_{0s}B^2_0}{\langle n\rangle B^2}\;\;,\;\;
	\delta N&=&\sum_s e_s\frac{\delta  {n}_s}{\langle n\rangle} \;\;,
	\delta J_\parallel=\sum_s \frac{\delta  {\bar{j}}_{\parallel,s}}{\langle n\rangle}\;\;, 
	\end{eqnarray}
	\revcolora{where $M_s=m_s/m_N$, $\bar{e}_s=e_s/e_N$ for the species `$s$',  $C_P=1/\bar\rho_N^2$, $C_A=\beta/\bar\rho_N^2$, $\bar\rho_N=\rho_N/R_N$,}  $\rho_N=m_Nv_N/(e_NB_0)$, $B_0$ in this work is chosen as the on-axis magnetic field, $\beta=2\mu_0 \langle n\rangle
	T_N/B_0^2$ and $\langle n\rangle$ is the volume averaged density. 
	
	The normalized equations of motion for the guiding center are expressed as follows,
	\begin{eqnarray}
	{\bm  {v}}_d &=& \frac{M_sB_0}{ \bar{e}_sB^2} {\rho}_N\left( {v}^2_\parallel+ {\mu} B\right){\bm b}\times\nabla B \;\;, \\
	\delta{\bm  {v}} &=& \frac{B_0}{B} {\rho}_N{\bm b}\times\nabla(\delta {\phi}- {v}_\parallel\delta  {A}_\parallel)\;\;, \\
	\dot{ {v}}_{\parallel0} &=& - {\mu}\partial_{\parallel}B\;\;,\\
	\label{eq:dvpardt_torus}
	\delta\dot{ {v}}_{\parallel} &=& -\frac{\bar{e}_s}{M_s}\left(\partial_\parallel\delta {\phi}+\partial_t\delta  {A}\right) \;\;.
	\end{eqnarray}
	The Poisson equation, the Amp\'ere's law and the guiding center's equations of motion in $(r,\phi,\theta)$ coordinates can be readily obtained (Appendix \ref{appendix:gcmotion}). 
	
	
	\subsection{The mixed PIC-PIF scheme using finite element and Fourier basis function}
	The field variables are decomposed using Fourier basis functions in $(\theta,\phi)$ directions and using finite elements in $r$ direction, 
	\begin{eqnarray}
	\delta\phi(r,\phi,\theta) &=& \sum_{n,m,k}\delta\phi_{nmk}\Lambda_k(r) e^{in\phi+im\theta}\;\;,\\
	\delta{A}_\parallel(r,\phi,\theta) &=& \sum_{n,m,k}\delta{A}_{\parallel,nmk}\Lambda_k(r) e^{in\phi+im\theta} \;\;, 
	\end{eqnarray}
	where $n$ and $m$ are the toroidal and poloidal harmonic numbers and $k$ serves as the radial index. 
	In the full $f$  scheme, the \revcolorb{physical} distribution function is represented by the markers,
	\begin{eqnarray}
	\label{eq:f_marker}
	f({\bm R},v_\parallel,\mu)& =& \frac{N_{ph}}{N_{ptot}} \sum_p\frac{w_p}{2\pi B_\parallel^*} \nonumber\\
	\delta({\bm R}&-&{\bm R}_p)\delta(v_\parallel-v_{\parallel,p})\delta(\mu-\mu_p)\;\;,
	\end{eqnarray}
	where $N_{ptot}$ is the marker number, $N_{ph}$ is the physical particle number, $w_p$ is set according to the initial physical and the marker distributions, $2\pi B_{||}^*$ is the Jacobian of the guiding center coordinates. The Poisson equation and the Amp\'ere's law are converted to the weak form,
\revcolora{
	\begin{eqnarray}
	\bar{\bar{M}}_{P,nn'mm'kk'} &\cdot&\delta\phi_{n'm'k'} = C_P\delta{N}_{nm}^k\;\;,  \\
	\bar{\bar{M}}_{A,nn'mm'kk'}&\cdot&\delta{A}_{\parallel,n'm'k'} = C_A\delta{J}_{nm}^k\;\;,  \\
	\bar{\bar{M}}_{P,nn'mm'kk'} &=& -\int_{r_0}^{r_1} dr \Lambda_{k}\nabla_{\perp,nm}\cdot g_{n-n',m-m'}\nabla_{\perp,n'm'}\ \Lambda_{k'}\;\;, \\
	\bar{\bar{M}}_{A,nn'mm'kk'} &=& -\int_{r_0}^{r_1} dr\Lambda_k \delta_{n-n',m-m'}\nabla_{\perp,nm}\cdot \nabla_{\perp,n'm'}\ \Lambda_{k'}\;\;, \\
	\label{eq:deltaN}
	\delta N_{nm}^k =& C_{p2g}\;&\sum_p \frac{R_0}{r_pR_p}w_p \Lambda_k(r_p)e^{-i(n\phi_p+m\theta_p)}\;\;,\\
	\label{eq:deltaJ}
	\delta J_{nm}^k =& C_{p2g}\;&\sum_p\frac{R_0}{r_pR_p} w_p v_{\parallel,p} \Lambda_k(r_p) e^{-i(n\phi_p+m\theta_p)}\;\;,
	\end{eqnarray}
}
	where $\nabla_{\perp,nm}$ is the Fourier representation of $\nabla_\perp$ with $\partial_\theta$ and $\partial_\phi$ replaced with $im$ and $in$ respectively, $C_{p2g}=(r^2_1-r^2_0)/(2N_{ptot})$, \revcolora{the Particle-in-Fourier method \cite{ameres2018stochastic,mitchell2019efficient,evstatiev2013variational}}
	\revcolor{ is adopted in the poloidal and toroidal directions, while the particle-in-cell is adopted in the radial direction, $\delta_{ij}=1$ for $i=j=0$, $\delta_{ij}=0$ for other $i,j$ values, $g=\sum_{n,m}g_{n,m}e^{in\phi+im\theta}$, and when calculating $\bar{\bar{M}}_{A/P,nn'mm'kk'}$ in the code, we  make use of the integration by parts}. 
	\revcolora{	In this work, we have adopted linear basis functions in the radial direction, in order to minimize the technical complexity, while the methods can be also applied with higher order basis functions in the future work. }
	\revcolorb{Equations \ref{eq:deltaN} and \ref{eq:deltaJ} are obtained from the velocity space integral of $f$ in Eq. \ref{eq:f_marker} and remain unchanged when  $B^*_\parallel\approx B$ is adopted. }
	Note that $\delta N_{nm}^k$ and $\delta J_{nm}^k$ are different from $\delta N_{nmk}$ and $\delta J_{nmk}$ defined by
	
	\begin{eqnarray}
	\delta{N}(r,\phi,\theta) &=& \sum_{n,m,k}\delta N_{nmk}\Lambda_k(r) e^{in\phi+im\theta}\;\;,\\
	\delta{J}(r,\phi,\theta) &=& \sum_{n,m,k}\delta J_{nmk}\Lambda_k(r) e^{in\phi+im\theta}\;\;.
	\end{eqnarray}
	
	\section{Implicit scheme with analytical treatment }
	\label{sec:numerical}
	In this section, for the sake of simplicity, we use the 1D problem to demonstrate the procedure of the implicit scheme and the analytical treatment. The key issue is to mitigate the numerical instability in the direction parallel to the magnetic field,  originating from $\partial_t \delta{A}_\parallel$ in the equation of motion, especially when the value of \revcolor{$\beta/(M_ek^2_\perp\rho_{N}^2)$} is large. The implicit scheme for the 3D tokamak geometry can be done with the same procedure, \revcolor{as briefly introduced in Section \ref{subsec:mixed_scheme}}. 
	
	\subsection{Shear Alfv\'en wave in uniform slab geometry}
	
	In the minimum model of SAW, the ion response is described with the polarization density, and only one kinetic species (electron) is kept. \revcolor{Noticing that $\bar{e}_s=-1$ for $s=e$}, the normalized equations are
	
	\begin{eqnarray}
	\label{eq:dldt1d}
		\frac{dl}{dt} &=& v_\parallel\;\;,\\
	\label{eq:dvdt1d}
		\frac{dv_\parallel}{dt} &=& \frac{1}{M_e}\left( \partial_\parallel\delta\phi+\partial_t\delta A_\parallel \right) \;\;, \\
	\label{eq:poisson1d}
		\nabla_\perp^2\delta\phi & = & C_P\delta N \;\;, \\
	\label{eq:ampere1d}
		\nabla^2_\perp\delta A_\parallel & = & C_A\delta J_\parallel \;\;. 
	\end{eqnarray}
	The Fourier \revcolor{components} of the density and current are obtained using particle-in-Fourier in the parallel direction,
	\begin{eqnarray}
	\label{eq:deltaNk}
		\delta{N}_{k_l} &=& \frac{1}{N_{ptot}}\sum_p e^{-ik_ll_p} \;\;, \\
	\label{eq:deltaJk}
		\delta{J}_{k_l} &=& \frac{1}{N_{ptot}}\sum_p v_{\parallel,p}e^{-ik_l l_p} \;\;,
	\end{eqnarray}
	where the Fourier decomposition is applied to the field and moment variables, e.g, $\delta N(l)=\sum_k \delta N_{k_l } \exp\{i{k_l}l\}$, and $k_l$ is the wave vector along $l$. 
	
	The energy conservation is tested for this 1D model in Section \ref{subsec:numerics_saw1d}. Using \revcolor{Eqs. \ref{eq:dldt1d}--\ref{eq:ampere1d}}, we have, \revcolor{theoretically, }
	\begin{eqnarray}
	\label{eq:energy_conserve}
	\revcolor{
		\frac{d}{dt} E_{tot}(t)}&=&0 \;\;, \\
	\label{eq:Etot_define}
		E_{tot}(t) &\equiv& 	E_{kin}(t) 
		+ E_B(t) +E_E(t) \;\;,\\
	\label{eq:energy_EB}
		E_E&=&\frac{k_\perp^2}{2C_P} \left|\delta\phi_k(t)\right|^2 
		\;\;,E_B = \frac{k_\perp^2}{2C_A}  \left|\delta A_k(t)\right|^2 \;\;,
	\end{eqnarray}
	where $	E_{kin}(t) $ is the particle kinetic energy. 
	\revcolora{Note that the energy conservation in the simulation also relies on the discretization scheme and the implicit treatment does not necessarily guarantee energy conservation. In this work, we use Eqs. \ref{eq:Etot_define}--\ref{eq:energy_EB} for the diagnosis to examine the quality of the scheme we adopted and the numerical implementation while the study of rigorous energy conserving schemes is out of the scope of this work.
	}
	
	\subsection{The implicit scheme for the particle-field system}
	\label{subsec:implicit_scheme}
	The implicit scheme is \revcolor{implemented} by applying the iteration scheme to the particle-field system. The purpose of the iteration between the particle pusher and the field solver is to achieve the implicit solution to the Crank-Nicolson scheme, i.e.,
	\begin{eqnarray}
	\label{eq:dldtCN}
	\frac{l^{t+\Delta t}-l^t}{\Delta t} &\equiv& \revcolor{\frac{\Delta l}{\Delta t} = } \frac{v_\parallel^{t+\Delta t}+v_\parallel^{t}}{2} \;\;, \\
	\label{eq:dvdtCN}
	\frac{v_\parallel^{t+\Delta t}-v_\parallel^t}{\Delta t} &\equiv &  \revcolor{\frac{\Delta v_\parallel}{\Delta t} =  } \frac{1}{2M_e}\partial_\parallel[\delta\phi^{t+\Delta t}+\delta\phi^{t}] 
	+\frac{1}{M_e} \frac{\delta{A}_\parallel^{t+\Delta t}-\delta{A}_\parallel^{t}}{\Delta t} \;\;,  \\
	\label{eq:poissonCN}
	\nabla_\perp^2\delta\phi^{t,t+\Delta t} & = & C_P\delta N^{t,t+\Delta t} \;\;, \\
	\label{eq:ampereCN}
	\nabla^2_\perp\delta A_\parallel^{t,t+\Delta t} & = & C_A\delta J_\parallel^{t,t+\Delta t} \;\;, 
	\end{eqnarray}
	\revcolor{where $\delta\phi$ and $\delta A_\parallel$ are taken at $l^t+\Delta l/2$ in Eq. \ref{eq:dvdtCN}}. In solving Eqs. \ref{eq:dldtCN} and \ref{eq:dvdtCN}, with the constraint $\Delta t v_\parallel k_\parallel\ll1$, it is applicable to take Taylor expansion \revcolora{ of the field perturbation $(\delta\phi,\delta A_\parallel)$, i.e., $\delta\phi(l^{t}+\Delta l/2)\approx\delta\phi(l^{t})+(\Delta l/2)\partial_l\delta\phi(l^t)$.  
    Then Eqs. \ref{eq:dldtCN} and \ref{eq:dvdtCN} yield the analytic expression as follows,} 
\revcolora{
	\begin{eqnarray}
	\label{eq:dldt_simple}
		\frac{\Delta l}{\Delta t} &=& \frac{1}{h(l^t)}\left[v_\parallel^t+W(l^t)  \right]\;\;,\\
	\label{eq:dvdt_simple}
		\frac{\Delta v_\parallel}{\Delta t} &=&
		\frac{2}{\Delta t}\left[\frac{\Delta l}{\Delta t}-v_\parallel^t\right] \nonumber\\
		&=&\frac{2}{h(l^t) \Delta t} \left\{
		W(l^t)+\frac{v_\parallel \Delta t}{2}\partial_lW(l^t)
		\right\}\;\;, \\
		h(l^t) &=& 1-\frac{\Delta t}{2}\partial_lW(l^t)\;\;,\\
		W(l^t) &=&\frac{1}{2M_e}
	\left\{
		\delta{A}_\parallel^{t+\Delta t}(l^t)
		-\delta{A}_\parallel^{t}(l^t)
		+\frac{\Delta t}{2}\partial_l
		\left[\delta\phi^{t+\Delta t}(l^t) +\delta\phi^{t}(l^t) \right]
\right\}\;\;.
	\end{eqnarray}
}
	A more rigorous way is to solve the nonlinear equations, i.e., Eqs \ref{eq:dldtCN} and \ref{eq:dvdtCN}, numerically, in order to achieve a good accuracy. In this work, Eqs. \ref{eq:dldtCN} and \ref{eq:dvdtCN} are solved by first defining the residual as follows,
\revcolora{
	\begin{eqnarray}
	\label{eq:R1particle}
	R_1&\equiv& l^{t+\Delta t}-l^t -\frac{\Delta t}{2} [v_\parallel^{t+\Delta t}+v_\parallel^t]\;\;,\\
	R_2&\equiv& v_\parallel^{t+\Delta t}-v_\parallel^t 
	-\frac{\Delta t}{2M_e} 
	\partial_l\left[
	\delta\phi^{t+\Delta t}(l^{t+\Delta t/2})+\delta\phi^{t}(l^{t+\Delta t/2})
	\right] \nonumber\\
	\label{eq:R2particle}
	&&-\frac{1}{M_e}[\delta A_\parallel^{t+\Delta t}(l^{t+\Delta t/2})-\delta A_\parallel^{t}(l^{t+\Delta t/2})]\;\;,
	\end{eqnarray} 
}
	and iterating $(l^{t+\Delta t},v_\parallel^{t+\Delta t})$ to reach  $R_1\rightarrow0, R_2\rightarrow0$. This can be achieved \revcolor{using the Newton iteration scheme,} by solving $\Delta l, \Delta v_\parallel$ as follows  
	\begin{eqnarray}
	\label{eq:deltalv_particle}
	\bar{\bar{M}}_R\cdot
	\begin{bmatrix}
	\Delta l \\ \Delta v_\parallel
	\end{bmatrix}
	&=&-
	\begin{bmatrix}
	R_1 \\ R_2
	\end{bmatrix}\;\;, \;\; 
	\bar{\bar{M}}_{R}=
	\begin{bmatrix}
	\frac{\partial R_1}{\partial l^{t+\Delta t}}\;\;,\;\;
	\frac{\partial R_1}{\partial v_\parallel^{t+\Delta t}}\\
	\frac{\partial R_2}{\partial l^{t+\Delta t}}\;\;,\;\;
	\frac{\partial R_2}{\partial v_\parallel^{t+\Delta t}}\\
	\end{bmatrix}\;\;, \nonumber\\
	\end{eqnarray}
	and by modifying $(l, v_\parallel)$ using $(\Delta l, \Delta v_\parallel)$ in the next particle iteration with the given $\delta\phi, \delta{A}_\parallel$. \revcolora{While Eqs. \ref{eq:dldt_simple} and \ref{eq:dvdt_simple} can serve as the initial condition of the rigorous calculation of the implicit particle solution, namely Eqs. \ref{eq:R1particle} -- \ref{eq:deltalv_particle}, the computational benefit is limited, since the particle iterative solver Eqs. \ref{eq:deltalv_particle} converges efficiently for small or moderate $k_\parallel v_\parallel\Delta t$. The more significant benefit of using Eqs. \ref{eq:dldt_simple} and \ref{eq:dvdt_simple} lies in using them as an approximate solution of Eqs. \ref{eq:dldtCN}--\ref{eq:dvdtCN}, without using the iterative particle solver at all, which gives almost the same results for the case in Fig. \ref{fig:saw_frequency_damping} as will be discussed in Section \ref{subsec:numerics_saw1d}. This serves as a tool for validation during the code development.}
	
	The main steps for iterations are as follows (all variables are at $t+\Delta t$ in the workflow, i.e., $\delta\phi=\delta\phi(t+\Delta t)$ etc),
	\begin{eqnarray}
	\label{wkflow:implicit}
	&&	\xrightarrow{\ref{iteration0}}
	\begin{Bmatrix}
	\delta\phi^{start}{} \\
	\delta{A}_\parallel^{start}{}
	\end{Bmatrix}^i
	\xrightarrow{\ref{iteration1}}
	\begin{Bmatrix}
	l{} \\
	v_\parallel{}
	\end{Bmatrix}^i
	\xrightarrow{\ref{iteration2}}
	\begin{Bmatrix}
	\delta{N}^{end}{} \\
	\delta{J}^{end}{}
	\end{Bmatrix}^i
	\nonumber\\
	&& 	\xrightarrow{\ref{iteration2}}
	\begin{Bmatrix}
	\delta\phi^{end}{} \\
	\delta{A_\parallel}^{end}{}
	\end{Bmatrix}^i
	\xrightarrow{\ref{iteration3}}
	\begin{Bmatrix}
	\delta\phi^{start}{} \\
	\delta{A}_\parallel^{start}{}
	\end{Bmatrix}^{i+1} 
	\end{eqnarray}
	\begin{enumerate}
		\item \label{iteration0}
		Each iteration starts with the given field $\{\delta\phi,\delta{A}\}^{start}(t+\Delta t)$.
		In each step from $t$ to $t+\Delta t$, as the first iteration ($i=1$), the explicit solution is used as the input of the first iteration. Namely, at time $t$, particles are pushed from $\{l(t)),v_\parallel(t)\}$ to $\{l(t+\Delta t),v_\parallel(t+\Delta t)\}$ using $\partial_\parallel\delta\phi(t)$ and $(\delta{A}_\parallel(t)-\delta{A}_\parallel(t-\Delta t))/\Delta t$. Then $\{\delta\phi(t+\Delta t), \delta{A}_\parallel(t+\Delta t)\}$ are calculated using $\{l(t+\Delta t),v_\parallel(t+\Delta t)\}$ by solving the Poisson equation and the Amp\'ere's law, and then serve as $\{\delta\phi^{start},\delta{A}_\parallel^{start}\}^{i=1}$. 
		\item \label{iteration1}
		Particles are pushed from $t$ to $t+\Delta t$ implicitly using $\{\delta\phi,\delta{A}\}^{start}(t+\Delta t)$ and $\{\delta\phi,\delta{A}\}(t)$ according to Eqs. \ref{eq:dldtCN} and \ref{eq:dvdtCN}, or, when $\Delta tv_\parallel k_\parallel\ll1$, to Eqs. \ref{eq:dldt_simple} and \ref{eq:dvdt_simple}. 
		\item \label{iteration2}
		In the end of the iteration,  $\{\delta\phi,\delta{A}\}^{end}(t+\Delta t)$ is calculated using Eqs. \ref{eq:poissonCN} and \ref{eq:ampereCN}. 
		\item \label{iteration3}
		The field perturbation for the next iteration is set according to    
		\begin{eqnarray}
		\begin{Bmatrix}
		\delta\phi(t+\Delta t)\\
		\delta{A}(t+\Delta t)
		\end{Bmatrix}^{start,i+1}
		=
		\begin{Bmatrix}
		\delta\phi(t+\Delta t)\\
		\delta{A}(t+\Delta t)
		\end{Bmatrix}^{start,i}
		+
		\begin{Bmatrix}
		\Delta\delta\phi\\
		\Delta\delta{A}
		\end{Bmatrix}, \nonumber
		\end{eqnarray}
		where $\Delta\delta\phi$ and $\Delta\delta{A}_\parallel$ are determined in such a way that in the $(i+1)$th iteration, 
		\begin{eqnarray}
		\label{eq:NJconverge}
		\begin{Bmatrix}
		\delta{N}^{start}(t+\Delta t) \\
		\delta{J}^{start}(t+\Delta t)
		\end{Bmatrix}^{i+1}
		=	    
		\begin{Bmatrix}
		\delta{N}^{end}(t+\Delta t) \\
		\delta{J}^{end}(t+\Delta t)
		\end{Bmatrix}^{i+1}\;\;,
		\end{eqnarray}
		or, at least, convergence occurs with respect to $i$. 
	\end{enumerate}
	Applying the Taylor expansion on the left hand side near $\{\delta\phi^{start},\delta{A}^{start}\}^i$, and the right hand side of Eq. \ref{eq:NJconverge} near $\{\delta{N}^{start}, \delta{J}^{start}\}^i$, we have
	\begin{eqnarray}
	\label{eq:corrM}
	\begin{Bmatrix}
	\begin{bmatrix}
	\frac{1}{C_P}\nabla_\perp^2 & 0\\
	0 & \frac{1}{C_A}\nabla_\perp^2 & 
	\end{bmatrix}
	-
	\bar{\bar{M}}_c 
	\end{Bmatrix}
	\cdot
	\begin{bmatrix}
	\Delta\delta\phi \\
	\Delta\delta{A}
	\end{bmatrix}
	=
	\begin{bmatrix}
	\Delta\delta{\tilde{N}} \\
	\Delta\delta{\tilde{J}}
	\end{bmatrix}\;\;,
	\end{eqnarray}
	where $\Delta\delta{\tilde{N}}\equiv\delta{N}^{end}-\delta{N}^{start}$, $\Delta\delta{\tilde{J}}\equiv\delta{J}^{end}-\delta{J}^{start}$, \revcolor{and the correction matrix is }
	\begin{eqnarray}
	\label{eq:Mc_theory}
	\bar{\bar{M}}_c\equiv 
	\begin{bmatrix}
	\frac{\partial\delta N^{t+\Delta t}}{\delta\phi^{t+\Delta t}}, \frac{\partial\delta N^{t+\Delta t}}{\delta{A^{t+\Delta t}}} \\
	\\
	\frac{\partial\delta J^{t+\Delta t}}{\delta\phi^{t+\Delta t}}, \frac{\partial\delta J^{t+\Delta t}}{\delta{A}^{t+\Delta t}} 
	\end{bmatrix}
	=
	\begin{bmatrix}
	\frac{k_\parallel^2(\Delta t)^2}{4M_e}, -\frac{ik_\parallel\Delta t}{2M_e} \\
	\\
	\frac{ik_\parallel\Delta t}{2M_e}, \frac{1}{M_e} 
	\end{bmatrix}\;\;.
	\end{eqnarray}
	The details of deriving the correction matrix \revcolor{$\bar{\bar{M_c}}$ in} Eq. \ref{eq:Mc_theory} are in Section \ref{subsec:mc_derivation}.	
	In summary, Eqs. \ref{eq:dldtCN}, \ref{eq:dvdtCN} (or \ref{eq:dldt_simple}, \ref{eq:dvdt_simple}),  \ref{eq:poissonCN}, \ref{eq:ampereCN}, \ref{eq:corrM} and \ref{eq:Mc_theory} embody our implicit scheme with analytical treatment and give the complete set for \revcolora{evolving} the system implicitly. 
	\revcolorb{All basic operations, such as the particle deposition, field scattering and field calculations in the implicit or the mixed implicit-explicit scheme (in Section \ref{subsec:mixed_scheme}) are similar to  those in the widely used explicit scheme even when the parallelization needs to be considered. The implicit particle solver treats each particle separately using given field information and can be parallelized easily. The  additional field equation (Eq. \ref{eq:Mc_theory}) is solved using the same way as the Poisson equation or the parallel Amp\'ere’s law, and thus can be parallelized easily as well. }

\revcolora{
	\subsection{The analytical correction matrix ($M_c$) of the implicit field solver (``moment enslavement'')}
	\label{subsec:mc_derivation}
	For obtaining the implicit solution to the field-particle system following the procedure \ref{wkflow:implicit}, the analytical correction matrix ($M_c$) of the implicit field solver in Eq. \ref{eq:Mc_theory} is derived, noticing that the moments $\delta N$ and $\delta J$ can be eventually written as functions of the fields $\delta\phi$ and $\delta A_\parallel$, which we refer to as ``moment enslavement''. 
	In deriving Eq. \ref{eq:Mc_theory}, firstly, the particle coordinates $(l,v_\parallel)$ at $t+\Delta t$ are functions of the fields at $t+\Delta t$, which follows the essence of the ``particle enslavement'' in a previous work \cite{chen2011energy}, 
	\begin{eqnarray}
	\label{eq:dldtCN2}
	l^{t+\Delta t}&=&l^t +  \frac{\Delta t}{2} (v_\parallel^{t+\Delta t}+v_\parallel^{t}) \;\;, \\
	\label{eq:dvdtCN2}
	v_\parallel^{t+\Delta t} &=& v_\parallel^t
	+ \frac{\Delta t}{2M_e}\partial_\parallel[\delta\phi^{t+\Delta t}+\delta\phi^{t}] 
	+\frac{\delta{A}_\parallel^{t+\Delta t}-\delta{A}_\parallel^{t}}{M_e} \;\;, 
	\end{eqnarray}
	which are from the Crank-Nicolson scheme Eqs. \ref{eq:dldtCN}, \ref{eq:dvdtCN}. Note the definition of the Fourier decomposition,
	\begin{eqnarray}
	\label{eq:fourierphi}
	\delta\phi(l)         &=&2Re[\delta\phi_{k_l}e^{ik_l l}]\;\;, \\
	\label{eq:fourierApar}
	\delta{A}_\parallel(l)&=&2Re[\delta{A}_{\parallel,k_l}e^{ik_l l}]\;\;,
	\end{eqnarray}
	where only the $\pm k_l$ Fourier components are kept for the sake of simplicity. Equations \ref{eq:dldtCN2}--\ref{eq:fourierApar} yield
	\begin{eqnarray}
	\frac{\partial l^{t+1}}{\partial\delta\phi^{t+\Delta t}_{k_l}} &=&\frac{ik_l\Delta t^2}{4M_e} e^{ik_l l^{t+\Delta t/2}} \;\;, \\
	\frac{\partial l^{t+1}}{\partial\delta{A}_{\parallel,k_l}^{t+\Delta t}} 
	&=&\frac{\Delta t^2}{2M_e} e^{ik_l l^{t+\Delta t/2}}   \\
	\frac{\partial v_\parallel^{t+1}}{\partial\delta\phi^{t+\Delta t}_{k_l}} 
	&=&\frac{ik_l\Delta t}{2M_e} e^{ik_l l^{t+\Delta t/2}}  \\
	\frac{\partial v_\parallel^{t+1}}{\partial\delta{A}_{\parallel,k_l}^{t+\Delta t}} 
	&=&\frac{1}{M_e} e^{ik_l l^{t+\Delta t/2}} 
	\end{eqnarray}
	
	Second, notice that the density and current perturbations are functions of particle coordinates $(l^{t+\Delta t},v_\parallel^{t+\Delta t})$,
	\begin{eqnarray}
	\label{eq:deltaNktp1}
	\delta{N}_{k_l}^{t+\Delta t} &=& \frac{1}{N_{ptot}}\sum_p e^{-ik_ll_p^{t+\Delta t}} \;\;, \\
	\label{eq:deltaJktp1}
	\delta{J}_{k_l}^{t+\Delta t} &=& \frac{1}{N_{ptot}}\sum_p v_\parallel^{t+\Delta t} e^{-ik_l l_p^{t+\Delta t}} \;\;,
	\end{eqnarray}
	which are equivalent to Eqs. \ref{eq:deltaNk} and \ref{eq:deltaJk} with $t+\Delta t$ explicitly written. 
	Then using Eqs. \ref{eq:dldtCN2}--\ref{eq:deltaJktp1} and chain rules, the correction matrix elements in Eq. \ref{eq:Mc_theory} are calculated as follows,
	\begin{eqnarray}
	\label{eq:dNdphi}
		\frac{\partial\delta N^{t+\Delta t}_{k_l}}{\partial\delta\phi^{t+\Delta t}_{k_l}} &=&
		\frac{1}{N_{ptot}} \sum_p\frac{\partial e^{-ik_ll_p^{t+\Delta t}} }{\partial\delta\phi^{t+\Delta t}_{k_l}}
		= \frac{k_l^2(\Delta t)^2}{4M_e}\frac{1}{N_{ptot}}\sum_pe^{-ik_l\Delta l_p/2}
		 \;\;,\\
	\label{eq:dNdapar}
		\frac{\partial\delta N^{t+\Delta t}_{k_l}}{\partial\delta{A}_{\parallel,k_l}^{t+\Delta t}} &=&
		\frac{1}{N_{ptot}} \sum_p\frac{\partial e^{-ik_ll_p^{t+\Delta t}} }{\partial\delta A_{\parallel,k_l}^{t+\Delta t}}
		= -\frac{ik_l \Delta t}{2M_e} \frac{1}{N_{ptot}}\sum_pe^{-ik_l\Delta l_p/2}
		\;\;,\\
	\label{eq:dAdphi}
		\frac{\partial\delta J^{t+\Delta t}_{k_l}}{\partial\delta\phi^{t+\Delta t}_{k_l}} &=&
		\frac{1}{N_{ptot}} \sum_p\frac{\partial v_\parallel e^{-ik_ll_p^{t+\Delta t}} }{\partial\delta\phi^{t+\Delta t}_{k_l}}
		= \frac{ik_l \Delta t}{2M_e }
		\frac{1}{N_{ptot}}\sum_p \left( 1-\frac{ik\Delta t v_{\parallel,p}}{2} \right)e^{-ik_l\Delta l_p/2} \;\;,	\\
	\label{eq:dAdapar}
		\frac{\partial\delta J^{t+\Delta t}_{k_l}}{\partial\delta A_{\parallel,k_l}^{t+\Delta t}} &=&
		\frac{1}{N_{ptot}} \sum_p\frac{\partial v_\parallel e^{-ik_ll_p^{t+\Delta t}} }{\partial\delta A_{\parallel,k_l}^{t+\Delta t}}
		= 
		\frac{1}{N_{ptot}}\sum_p \left( 1-\frac{ik\Delta t v_{\parallel,p}}{2} \right)e^{-ik_l\Delta l_p/2} \;\;,
	\end{eqnarray}
	where $\sum_p e^{-ik_l\Delta l_p/2}\approx N_{ptot}$ can be used when $k_l\Delta l_p/2\ll1$. Another time discretization with the fields solved at $t+\Delta t/2$ but particles pushed along $t,\;t+\Delta t,\ldots$ can eliminate the $e^{-ik_l\Delta l_p/2}$ factor and will be studied in the future. For the shear Alfv\'en wave studied in this work, the $v_{\parallel,p}$ terms in the bracket are ignored since the equilibrium flow is zero and the perturbed fluid velocity (normalized to thermal velocity) is infinitesimal. Then, the correction matrix can be obtained as shown in Eq. \ref{eq:Mc_theory} without explicitly specifying the Fourier mode number in the subscript and with $k_l$ replaced by $k_\parallel$. 
	
	By using the analytical results in Eqs. \ref{eq:dNdphi}--\ref{eq:dAdapar} or Eq. \ref{eq:Mc_theory}, the numerical calculation of $M_c$ can be avoided. On the one hand, the analytical solution gives the accurate solution of $M_c$ while the numerical calculation of $M_c$ relies on the convergence of the derivative calculation of $\delta N_{k_l}$ and $\delta J_{k_l}$ with respect to the variation of $\delta\phi_{k_l}$ and $\delta A_{\parallel,k_l}$. On the other hand, in calculating $M_c$ analytically, no operation (such as particle push) on each single particle is needed  but only the fluid-like terms $\sum_p e^{-ik_l\Delta l_p/2}$ and $\sum_p v_{\parallel,p} e^{-ik_l\Delta l_p/2}$ are needed, which can be simplified further in the small perturbation limit, as adopted in Eq. \ref{eq:Mc_theory}. 
}

	\subsection{The mixed implicit-explicit scheme}
	\label{subsec:mixed_scheme}
	While the above implicit scheme is based on the 1D model, the implicit scheme in \revcolora{tokamak} plasmas can be implemented by applying either fully 3D implicit scheme on the same footing, or, as adopted in this work, the mixed implicit-explicit scheme, \revcolor{inspired by the theoretical mixed WKB-full-wave approach \cite{zonca1993theory,lu2012theoretical,lu2013mixed}}. Using this mixed scheme, only the \revcolora{fast} parallel motion terms are treated implicitly but the other terms are treated using an explicit scheme, such as the Runge-Kutta method, as adopted in this work. 
\revcolora{
	The splitting of the guiding center's equations of motion (Eqs. \ref{eq:dRdt} and \ref{eq:dvpardt}) are as follows,
	\begin{eqnarray}
		\label{eq:dxvdt_emp}
		\frac{d{\bm R}^{E}}{dt} &=& {\bm v}_d+\delta{\bm v} \;\;,
		\;\;
		\frac{dv_\parallel^E}{dt} =  \dot{v}_{\parallel0}\;\;,\\
		\frac{d{\bm R}^{I}}{dt} &=& {\bm v}_\parallel \;\;,
		\;\;
		\frac{dv_\parallel^I}{dt} = \delta\dot{ {v}}_{\parallel} \;\;,
	\end{eqnarray}
	where ${\bm v}_d$, $\delta{\bm v}$, $\dot v_{\parallel 0}$ and $\delta\dot{v}_\parallel$ are defined in Eqs. \ref{eq:vd}--\ref{eq:ddeltavpardt}. In each sub step of the Runge-Kutta step, as the first operation, the explicit increment $(\Delta{\bm R}^E, \Delta v_\parallel^E)$ is calculated according to Eq. \ref{eq:dxvdt_emp}. 
	Then the increment $(\Delta{\bm R}^I, \Delta v_\parallel^I)$ is calculated using the implicit scheme in a  similar way in Eqs. \ref{eq:dldtCN}--\ref{eq:ampereCN} with $(\Delta{\bm R}^E, \Delta v_\parallel^E)$ included in $({\bm R},v_\parallel)^{t+\Delta t}$ and fixed as constants when solving for the implicit solution, 
	\begin{eqnarray}
	\label{eq:dldtCN_mix}
		&&\frac{\Delta{\bm R}^I}{\Delta t} = \frac{\bm{v}_\parallel^{t}+(\bm{v}_\parallel^{t}+\Delta \bm{v}_\parallel^E+\Delta \bm{v}_\parallel^I)}{2} \;\;, \\
	\label{eq:dvdtCN_mix}
		&&\frac{\Delta v_\parallel^I}{\Delta t} = -\frac{\bar{e}_s}{2M_e}\partial_\parallel[\delta\phi^{t+\Delta t}+\delta\phi^{t}] 
		-\frac{\bar{e}_s}{M_e} \frac{\delta{A}_\parallel^{t+\Delta t}-\delta{A}_\parallel^{t}}{\Delta t} \;\;, \nonumber \\ \\
	\label{eq:poissonCN_mix}
		&&\nabla_\perp\cdot g\nabla_\perp\delta\phi^{t,t+\Delta t} = C_P\delta N^{t,t+\Delta t} \;\;, \\
	\label{eq:ampereCN_mix}
		&&\nabla^2_\perp\delta A_\parallel^{t,t+\Delta t} = C_A\delta J_\parallel^{t,t+\Delta t} \;\;, 
	\end{eqnarray}
	where $\bm{v}_\parallel=v_\parallel {\bm   b}$, $g$ is defined in Eq. \ref{eq:definegNJ}, $\delta\phi^{t+\Delta t}$, $\delta A_\parallel^{t+\Delta t}$, $\delta N^{t+\Delta t}$ and $\delta J_\parallel^{t+\Delta t}$ are evaluated using the particle information at $t+\Delta t$, i.e., $(\bm{R}_\parallel^{t+\Delta t},v_\parallel^{t+\Delta t})=({\bm{R}}_\parallel^t+\Delta {\bm R}_\parallel^E+\Delta {\bm R}_\parallel^I,v_\parallel^t+\Delta v_\parallel^E+\Delta v_\parallel^I)$. The implicit particle-field solver is implemented following the workflow in Eq. \ref{wkflow:implicit} for the 1D case. The particle's implicit solution with given fields and $(\Delta{\bm R}^E,\Delta v^E_\parallel)$ is obtained following Eqs. \ref{eq:R1particle}--\ref{eq:deltalv_particle}. The correction to the field is obtained in the same way as shown in Eq. \ref{eq:corrM}, in order to achieve the implicit field-particle solution. 
}
	
	\section{Numerical results}
	\label{sec:result}
	
	The one dimension SAW model is implemented in Matlab and the electromagnetic model for tokamak plasmas is implemented in Fortran. In this section, the simulation results are presented for these two cases. For the simulation in tokamak plasmas, the EP driven TAE case defined by the ITPA group is adopted \cite{konies2018benchmark}. LIGKA is run for the calculation of the TAE eigenvalue \cite{lauber2007ligka}, and for the comparison with the particle simulation results. 
	
	\subsection{Shear Alfv\'en wave in 1D uniform plasma}
	\label{subsec:numerics_saw1d}
	As the benchmark of the particle simulation using the implicit scheme in 1D geometry (Eqs. \ref{eq:dldt1d}--\ref{eq:ampere1d}), the electromagnetic dispersion relation in uniform plasma is adopted as the analytical solution, \cite{kleiber2016explicit}
	\begin{eqnarray}
	\label{eq:disp_saw}
	D&=&1-\frac{2\beta[1+\bar{\omega}Z(\bar{\omega})]}{M_e(k_{\perp}\rho_{ti})^2} 
	\left(
	\bar{\omega}^2-\frac{M_e}{\beta}
	\right)
	=0\;\;, \nonumber\\
	\end{eqnarray}
	where $k_\perp$ is the perpendicular wave number, $Z$ is the plasma dispersion function, $\bar{\omega}=\omega/\omega_{te}$, $\omega_{te}=v_{te}k_\parallel$. 
	
	The simulation parameters are as follows. 
	\revcolora{The particle-in-Fourier scheme has been used with one harmonic ($e^{\pm ik_l l}$) in the direction parallel to the magnetic field.} $k_\perp\rho_N=0.2$, $\beta/M_e$ is chosen in the range of $[1/16,32]$ in the parameter scan, which covers the typical regime of tokamak plasmas, e.g., $\beta=1\%$, $M_e=1/1836$, i.e., $\beta/M_e=18.36$. The roots of the SAW are calculated in the complex space, by solving Eq.  \ref{eq:disp_saw}. The least damped roots with $\bar{\omega}=\pm0.319-0.0017428i$ ($\beta/M_e=10$) correspond to the SAW and serves as the analytical solution for the comparison with our particle simulation, while the other heavily damped roots can be hardly observed in the particle simulations.
	
	The particle simulation based on the implicit scheme shows its performance in SAW studies, as shown in Fig. \ref{fig:conservation_energy}. In this case, the marker number \revcolor{$N_{ptot}=10^6$,} the time step $\Delta t=0.01\cdot T_{SAW}$, where the SAW period $T_{SAW}=2\pi/(v_Ak_\parallel)$, $\beta/M_e=4$. 
	\revcolor{The simulation results in $15T_{SAW}$ show that the Landau damping of the initial perturbation occurs during $t\in[0,5T_{SAW}]$ and after that, the wave-particle nonlinear interaction leads to the energy transfer between the wave and particles back and forth.  } The total particle kinetic energy and the wave energy are calculated as shown in the top frame. 
	As the wave gets damped, the total particle kinetic energy grows, \revcolor{and vice versa}. In the middle frame, the $\delta B$ component (magenta line) and the $\delta E$ component (blue line)  oscillate with the same amplitude, but with 90 degrees of phase shift. The total energy (blue line in the first row) indicates good conservation properties. 
	\revcolorb{The relative error of the total energy is shown in the bottom frame, demonstrating that $E_{tot}(t)/E_{tot0}-1$ increases to $\sim1\%$ in $3T_{SAW}$ and after that, stays in the magnitude lower than $2\%$, where $E_{tot0}$ is the initial total energy (longer time simulation will be shown in Fig. \ref{fig:converge_energy}). It can be shown that this artificial energy loss is small compared with the theoretical wave damping rate, i.e.,  $\gamma_{artificial}\approx2.6\%\gamma_{theory}$, where $\gamma_{theory}/\omega_{TAE}=0.01009$ from Eq. \ref{eq:disp_saw}. The energy conservation can be improved efficiently as the step size $\Delta t$ is reduced, as shown in Fig. \ref{fig:converge_energy}. The SAW is simulated in $100T_{SAW}$ and the relative error of the total energy is shown in the top frame. The relative error stays on a steady level during the nonlinear phase, as shown in the top frame. As shown in the bottom frame, the relative error is reduced significantly as $\Delta t$ decreases. As $\Delta t/T_{SAW}$ is reduced from $0.01$ to $0.005$, the absolute value of the average relative error of the total energy decreases from $1.58\%$ to $0.41\%$ for $N_{ptot}=10^6$. The relative error is not sensitive to the marker number $N_{ptot}$ in the range of $N_{ptot}=5\cdot10^5,10^6,2\cdot10^6$. For small damping cases in the large $\beta/M_e$ limit or small $\beta/M_e$ limit, the relative error is significantly smaller and the error in calculating the damping rate is also under control. 
	}
	
	The real frequency and the damping rate of the SAW calculated using the implicit particle code and the eigenvalue solver (Eq. \ref{eq:disp_saw}) are shown in Fig. \ref{fig:saw_frequency_damping}. The marker number is $10^6$. For the weakly damped SAW (e.g., $\beta/M_e=16$), the frequency $\omega_r$ and the damping rate $\gamma$ are fitted in $10\cdot T_{SAW}$, while for the SAW with larger damping rate (e.g., $\beta/M_e=1/2$), $\omega_r$ and $\gamma$ are fitted in $4\cdot T_{SAW}$. 
	\revcolora{When choosing the time step size $\Delta t$, the limit due to A) the numerical stability, B) the accuracy and C) the convergence is considered. First, $\Delta t$ needs to be smaller than a critical value to avoid numerical instabilities. Since the implicit solver takes the explicit trial solution as the starting point, as clarified in the Step 1 of Eq. \ref{wkflow:implicit}, $\Delta t$ \mbox{can not} be too large so that the implicit solver can find the physical implicit solution near the explicit solution. For $\beta/M_e=1/16$, numerical instability (crash) appears as $\Delta t$ increases from $1.0T_{SAW}$ to $1.5T_{SAW}$, but the simulation is crash-free for $\Delta t/T_{SAW}<=1$. As $\beta/M_e$ increases, the maximum $\Delta t$ needed to avoid numerical instabilities drops. Second, in order to fit the frequency and the damping rate accurately, we have to use at least 20 points in one wave period. Third, $\Delta t$ is small enough so that reasonable convergence can be observed as $\Delta t$ is varied. Specifically, the } maximum time step size used in the scan is $\Delta t=T_{SAW}/20$ for $\beta/M_e=1/16$ and the minimum one is $\Delta t=T_{SAW}/120$ for  $\beta/M_e=32$.  
	\revcolorb{In the small $\beta/M_e$ limit, $|\partial_t\delta A_\parallel|\ll|\partial_{\parallel}\delta\phi|$ and $\delta E_\parallel$ is mainly contributed by the electrostatic scalar potential $\partial_\parallel\delta\phi$ while in the large $\beta/M_e$ limit, $\partial_t\delta A_\parallel\approx-\partial_{\parallel}\delta\phi$ and $|\delta E_\parallel/|\partial_{\parallel}\delta\phi|\ll1$, as can be found from Eq. \ref{eq:disp_saw}. As a result, when $\beta/M_e$ increases, $\Delta t$ needs to be smaller in order to treat the $\partial_t\delta A_\parallel$ term and its cancellation with $\partial_\parallel\delta\phi$ properly. }
	The implicit scheme shows its capability in the small electron mass condition, which is usually a challenge in kinetic particle simulations, due to the quick electron response to $\delta E_\parallel$.
	The scan with fixed $\beta$ (but varying $m_e$) and that with fixed $m_e$ (but varying $\beta$) show no difference in the mode eigenvalue, which is obvious from the dependence of the analytical dispersion relation Eq. \ref{eq:disp_saw} on $\beta/M_e$. 
	
	\subsection{Toroidicity induced Alfv\'en eigenmode damping and excitation \revcolorb{in three dimensional axisymmetric tokamak}}
	To simulate the Alfv\'en modes in tokamak plasmas, Eqs \ref{eq:poisson_torus}--\ref{eq:dvpardt_torus} are solved using the implicit particle scheme. The TAE is simulated using the parameters of the widely studied ITPA case \cite{konies2018benchmark}. The major radius $R_0=10\;\text{m}$, minor radius $a=1\;\text{m}$, on-axis magnetic field $B_0=3\;\text{T}$, the safety factor profile $q(r)=1.71+0.16r^2$. The electron density is constant with $n_{e0}=2.0\cdot10^{19}\;\text{m}^{-3}$, $T_e=1\;keV$. The EP density profile is
	\begin{eqnarray}
	n_f(r)=n_{f0}c_3 \exp\left( -\frac{c_2}{c_1} \tanh\frac{r-c_0}{c_2}\right)\;\;,
	\end{eqnarray}
	where $n_{f0} =1.44131\cdot10^{17}\; \text{m}^{-3}$, the subscript `$f$' indicates EPs (fast particles), $c_0 = 0.491 23$, $c_1 =0.298 228$, $c_2 =0.198 739$, $c_3 =0.521 298$. The EP temperature is $400\;keV$. Since the dominant bulk ion response is already included in the polarization density in the Poisson equation, 
	\revcolorb{only kinetic electrons and fast ions but no kinetic bulk ions} are included in this work.  
	
	\subsubsection{Numerical verification}
	The field solver is tested using the Method of Manufactured Solutions (MMS), without including particles. The Poisson solver and the Amp\'ere solver are both constructed from the mass and stiffness matrices, corresponding to $\partial^2/\partial r^2$, $\partial/\partial r$ and $f(r)$, where $f(r)$ is a function of $r$. As a result, testing the Amp\'ere solver is sufficient for the numerical verification of the basic field operators. The Amp\'ere's law can be written as (Eq. \ref{eq:ampere_in_rtp})
	\begin{eqnarray}
	\left(\frac{\partial^2}{\partial r^2}+\frac{1}{r}\frac{\partial}{\partial r}-\frac{m^2}{r^2}\right)\delta{A}_{\parallel,m} =  C_A \delta J_{m}\;\;,
	\end{eqnarray}
	where the toroidal mode number $n$ is omitted 
	\revcolorb{in the subscript since $n=-6$ is fixed in this whole section, and the perpendicular Laplacian operator in Eq. \ref{eq:ampere_torus} is replaced with that in the poloidal plane by ignoring the terms smaller by a factor of $r^2/(qR)^2$}. 
	The analytical solution is given as
	\begin{eqnarray}
	\label{eq:mms_sol}
	\delta{A}_{\parallel,m,ana} &=& c_0+c_1 r+a_J J_m(r) + e^{-\left(\frac{r-r_c}{W}\right)^2} \;\;, \\
	\label{eq:mms_rhs}
	C_A\delta J_{m,ana}  &=& a_2r^2+a_3r^3+a_+r^m+a_-r^{-m} -a_JJ_m(r) \nonumber\\
	+e^{-\left(\frac{r-r_c}{W}\right)^2}&\times&\left[ \frac{4(r-r_c)^2}{W^4} -\frac{2}{W^2} -\frac{2(r-r_c)}{rW^2} -\frac{m^2}{r^2} \right]\;\;, \nonumber\\
	\end{eqnarray}
	\revcolor{where $J_m$ is the Bessel function.} The right hand side of the Amp\'ere's law is set to Eq. \ref{eq:mms_rhs} and the numerical solution $\delta{A}_{\parallel,m,num}$ is compared with $\delta{A}_{\parallel,m,ana}$ in Eq. \ref{eq:mms_sol}. 
	The relative error in the numerical solution $\sqrt{[\sum_k(f_{num,k}-f_{ana}(r_k))^2]/\sum_k f^2_{ana}(r_k)}$, where $k$ indicates the radial grid index, $f=\delta{A}_{\parallel,m}$, is shown in Fig. \ref{fig:mms_test}, where $N_r$ is the radial grid number. Reasonable convergence of the field solver is observed. In our simulation, by choosing $N_r=60$, the relative error in $\delta{A}_\parallel$ for given $\delta{J}$ is at the level of $10^{-3}$ in the field solver.
	
	The particle pusher is tested by the diagnosis of the particle trajectory and the two constants of motion, namely, the energy $E$ and the canonical toroidal momentum $P_\phi$. 
	\revcolor{The particle trajectories are shown in Fig. \ref{fig:particle_EP_error}. }
	The particle temperature is $T_f=400\;keV$, \revcolor{the on-axis magnetic $B_0$ is $3\;T$, the time step  $\Delta t=0.025T_{f,trans}$, the transit period $T_{f,trans}=2\pi q_{r=0.5a}R_0/\sqrt{2T_f/m_f}$. The particle trajectories in $100 T_{f,trans}$ are calculated. }
	For passing particles (the upper row), $\mu=0.04v_{ts}^2/B_0$, $v_\parallel\in[-2v_{ts},2v_{ts}]$ at $r=0.5$, $\theta=0$. For trapped 
	particles (the lower row), $\mu\in[0.15v_{ts}^2,v_{ts}^2]$, $v_\parallel=-0.2v_{ts}$ at $r=0.8$, $\theta=0$. 
	\revcolor{The corresponding  root-mean-square relative errors in $E$ and $P_\phi$ are lower than $5\cdot10^{-5}$ for all particles in $100 T_{f,transit}$.
	}

	In order to test the convergence of the implicit field-particle solver, the relative correction in $\delta\phi$ and $\delta{A}_\parallel$ in every iteration are analyzed. In the iteration procedure Eq. \ref{wkflow:implicit}, the iteration can be terminated when $Err(\delta\phi)\equiv\sqrt{\sum(\Delta\delta\phi)^2/\sum\delta \phi^2}$ and $Err(\delta{A}_\parallel)\equiv\sqrt{\sum(\Delta\delta{A}_\parallel)^2/\sum\delta{A}_\parallel^2}$ are small enough (typically, $10^{-8}$).  The convergence of the implicit particle-field solver in a typical simulation is shown in Fig. \ref{fig:convergence}. Two time slices are selected for the diagnosis of the convergence. In 15 iterations, the relative error in $\delta\phi$ and $\delta{A}_\parallel$ decreases to $10^{-8}$ and lower, as a good indication of convergence.  
	
	\subsubsection{Toroidicity induced Alfv\'en eigenmode w/o EPs}
	The TAE is simulated with no EPs applied firstly. 
	Two cases of the TAE damping are studied. In the first case, we choose $m_i/me=200$, since this is the parameter used in the EP driven TAE in the next section and previous ORB5 simulations \cite{biancalani2017nonlinear}. In the second case, we choose $m_i/m_e=1836$, in order to compare with the previous results where $m_i/m_e=1836$ is used for the calculation of the TAE damping \cite{konies2018benchmark}. LIGKA is run for both cases as the benchmark. 
	The initial density perturbation \revcolorb{with the amplitude of $\delta n(r=0.5a)/n_{e0}=4\cdot10^{-3}$} is loaded by initializing markers' displacement. The initial density perturbation has a Gaussian shape $\delta N_p(r)=\sigma_p \exp\{-(r-r_{pc})^2/W_p^2\}$. 
	Since the noise level in density is estimated as $\sigma_{noise}=1/\sqrt{N_{ptot}/N_r}$, the amplitude of the initial density perturbation is set to at least $2$ times of $\sigma$, i.e., $\sigma_p=2\sigma_{noise}$ in order to simulate the TAE mode structure and the time evolution clearly. The Gaussian shape $\exp\{-(r-r_{pc   })^2/W_p^2\}$ of the density perturbation is set to be as close as possible to the TAE eigenmode \revcolorb{with the $m=10,11$ poloidal harmonics as the dominant ones near $r=0.5a$}. In practice, we adopted $W_p=0.025$, $r_{pc}(m=10)=0.47$, $r_{pc}(m=11)=0.51$. The marker number is $N_{ptot}=16\cdot10^6$, the time step size is $\Delta t=T_{TAE}/100$ for $m_i/m_e=200$ and $\Delta t=T_{TAE}/800$ for $m_i/m_e=1836$. The simulation completes $10\;T_{TAE}$ on 8 computing nodes within around 10 hours for the $m_i/m_e=200$ case and $12.5\;T_{TAE}$ within around 80 hours for the $m_i/m_e=1836$ case, with each node containing two Intel Xeon Gold 6148 processors (Skylake (SKL), 20 cores @ 2.4 GHz).
	
	The time evolution of the TAE for $m_i/m_e=200$ is shown in the top left frame of Fig. \ref{fig:tae_t1d_mode2d}. The physics value of the electrostatic potential perturbation, $\delta\phi_c$,  is measured at $r=0.48,\theta=0$. The time evolution is clear, indicating the proper simulation of the TAE. The anatyical TAE frequency $\omega_{TAE}=v_A/(2qR_0)=417.8\cdot10^3rad/s$ is used as the reference. The real frequency fitted during $t/T_{TAE}\in[1,10]$ gives the real frequency $\omega_r/\omega_{TAE}=0.9615$, i.e., $\omega_r=401.7\cdot10^3rad/s$. The damping rate from the simulation is $\gamma/\omega_{TAE}=-0.011999$, i.e., $\gamma=-5013/s$.  
	\revcolor{As a study regarding the sensitivity of the initial density perturbation, we ran the    case with $\delta n_{r=0.5a}/n_{e0}=8\cdot10^{-3}$ (keeping other parameters unchanged), and the damping rate is slightly different (by $\sim2.5\%$) compared with the one with $\delta n_{r=0.5a}/n_{e0}=4\cdot10^{-3}$. 
	}
	The mode frequency and the damping rate are compared with the results from LIGKA \cite{lauber2007ligka}. 
	LIGKA computes the complex eigenvalue of the linearized gyrokinetic equations using numerically computed unperturbed orbit integrals for both electrons and ions. 
	The value from this LIGKA numerical model $\gamma/\omega_r=-1.293\%$ is close to the TRIMEG-GKX result $\gamma/\omega_r=-1.248\%$. 
	For the $m_i/m_e=1836$ case, the frequency and the damping rate are $(\omega_r,\gamma)=(0.98142,-0.004907)\cdot\omega_{TAE}$ by using the wave energy $E_E$ defined in Eq. \ref{eq:energy_EB} during $t/T_{TAE}\in[5,12.5]$, in order to enhance the signal for this weakly damped mode. 
	\revcolorb{Here the wave energy integral in the whole plasma ($E_E$) is calculated during $t/T_{TAE}\in[5,12.5]$, during which the mode structure is stable, and the linear decay of $log(E_E)$ is  clear. As a result, the fitted damping rate and the frequency of $\sqrt{E_E}$ represents those of the TAE. The obtained $\gamma$ and $\omega$ can be viewed as the average value at different radial locations using the scalar potential, which gives a good estimate for this weakly damped case.} The value from LIGKA  ($\gamma/\omega_r=-0.5008\%$) is close to the TRIMEG-GKX result ($\gamma/\omega_r=-0.5000\%$) for the realistic electron mass ratio.
In the previous benchmark results \cite{konies2018benchmark}, using the realistic electron mass, the damping rate is $-1103/s$ for GYGLES, $-567/s$ (co propagating TAE) or $-1705$ (counter propagating TAE) for EUTERPE. In recent ORB5 simulation, the damping rate is $1825/s\sim2190/s$ (Fig. 6 of  Ref. [\!\!\citenum{vannini2020gyrokinetic}]). In our simulation, both co- and counter-propagating TAEs are included and the estimated damping $\gamma=2050/s$ is also comparable to other codes.
		
	The 2D TAE mode structures are shown in the top middle and top right columns of Fig. \ref{fig:tae_t1d_mode2d}.  The mode width is consistent with previous simulation results with full width at half maximum $\Delta r\approx0.06$ in the mode envelope. The magnitude of the $m=10$ poloidal harmonic is larger than those of other harmonics, which is consistent with the observations by other codes such as LIGKA, GYGLES, ORB5 and EUTERPE  \cite{konies2018benchmark}. 
	
	\subsubsection{Energetic particle driven Toroidicity induced Alfv\'en eigenmode}
	\label{subsec:EP_TAE}
	For the EP driven TAE, the marker numbers for electrons and EPs are $N_{ptot,e}=128\cdot10^6$, $N_{ptot,f}=32\cdot10^6$, and the time step size is $\Delta t=T_{TAE}/100$. 
	\revcolorb{The initial density perturbation with the amplitude of $\delta n(r=0.5a)/n_{e0}=5\cdot10^{-4}$ is loaded by initializing markers' displacement. For the corresponding $\delta\phi$, the $m=10,11$ TAE component is not dominant compared with other components ($m=8,9,12,13$) and serves as a seed for the EP driven TAE.}
	The simulation completes on 24 computing nodes within around 36 hours. 
	The time evolution of the EP driven TAE is shown in the bottom left frame of Fig. \ref{fig:tae_t1d_mode2d}. Since the initial perturbation \revcolor{(especially the $m=8,9,12,13$ perturbation) is significantly different than the EP driven TAE, it is} damped firstly during $0<t/T_{TAE}<2$ and then the TAE is excited by EPs. 
	The real frequency fitted during $t/T_{TAE}\in[4,10]$ is $\omega_r/\omega_{TAE}=0.9276$. The growth rate fitted during the growing phase ($2.5<t/T_{TAE}<5.5$) gives $\gamma/\omega_{TAE}=0.090806$ (most codes give $\gamma/\omega_{TAE}\approx9\%\sim12\%$ \cite{konies2018benchmark}).  
	
	The 2D mode structure and the radial profile of the poloidal harmonics at $t/T_{TAE}=5.5$ are shown in the bottom middle and bottom right of Fig. \ref{fig:tae_t1d_mode2d}. The broadening of the radial envelope (full width at half maximum $\sim0.12$ from the bottom right frame) is larger by $100\%$ than that of the TAE damping case in the top right frame. This is due to the EPs' non-perturbative effects \revcolor{on} broadening the mode structure \cite{wang2013radial,biancalani2017nonlinear}. Another feature is the mode structure symmetry breaking, namely, the mode structure distortion, due to the EPs' contribution to the non-Hermitian part of the dispersion relation \cite{ma2015global,lu2018kinetic,lu2019theoretical}. More quantitative studies on the properties of the mode structure symmetry breaking using this full $f$  simulation and its effects on the EP transport  \cite{meng2020effects}, will be performed in future work. 
	

	
	\begin{figure}\centering
		\includegraphics[width=0.45\textwidth]{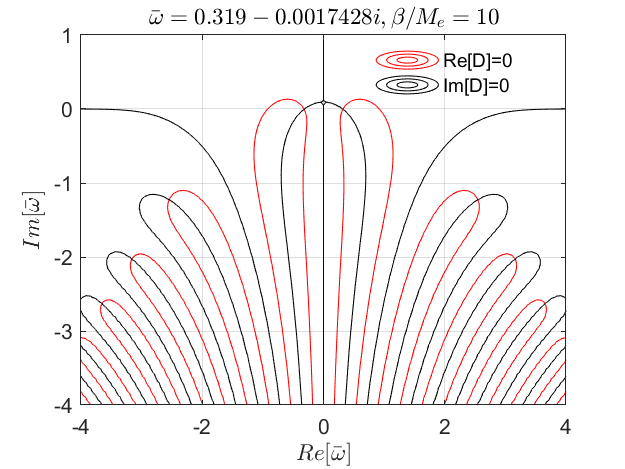}	
		\caption{The roots of the dispersion relation. The red or the black lines indicate the real or the imaginary parts of the SAW dispersion relation, Eq. \ref{eq:disp_saw}, and their intersection gives the eigenmode solution $D(\bar{\omega})=0$. The least damped root with maximum $|Im(\zeta)|$ corresponds to the SAW. 
		}	\label{fig:roots2d}
	\end{figure}
	
	\begin{figure}\centering
		\includegraphics[width=0.45\textwidth]{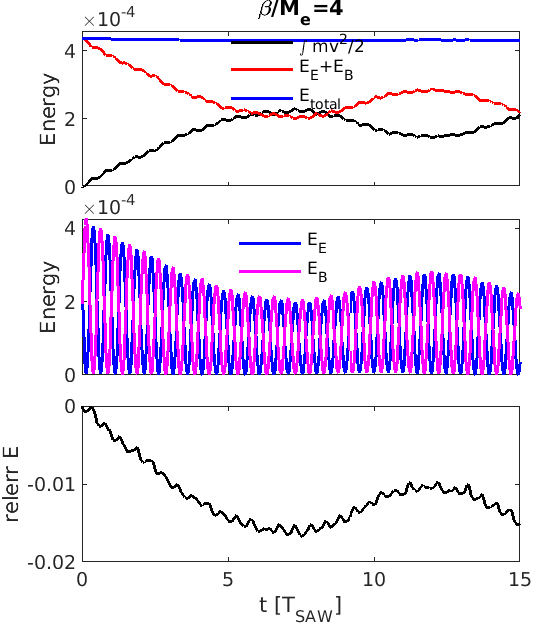}
		\caption{Top: time evolution of the particle kinetic energy (black line), wave energy (red) and the total energy (blue); middle: the wave energy $E_E$ and $E_B$ defined by Eq. \ref{eq:energy_EB}; 
		\revcolora{bottom: relative error of total energy, defined as $E_{tot}(t)/E_{tot0}-1$, where $E_{tot0}$ is the initial total value.} 
		}
		\label{fig:conservation_energy}
	\end{figure}
	
	\begin{figure}\centering
	\includegraphics[width=0.45\textwidth]{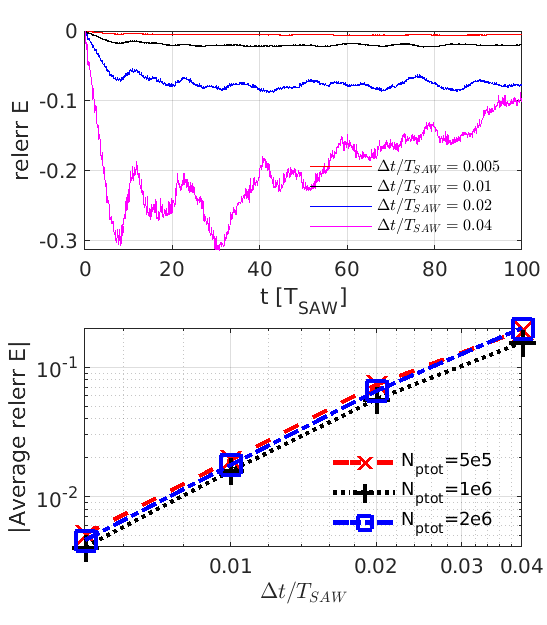}
	\caption{
	\revcolor{Top: the relative error of the total energy for different time step size $\Delta t$ and fixed marker number $N_{ptot}=10^6$. Bottom: the time-averaged absolute value of the relative error for different $\Delta t$ and marker number.
	}
	} 
	\label{fig:converge_energy}
	\end{figure}
	
	\begin{figure}\centering
		\includegraphics[width=0.48\textwidth]{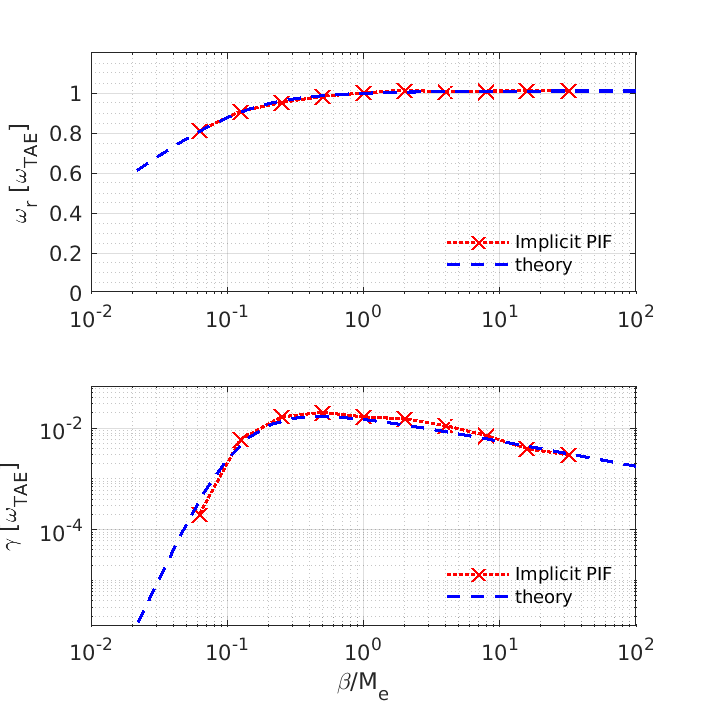}
		\caption{The theoretical value solved from Eq. \ref{eq:disp_saw} (blue broken lines) and the simulation results using the implicit particle method (crosses) of the real frequency (top) and the damping rate (bottom) of the SAW. 
		}
		\label{fig:saw_frequency_damping}
	\end{figure}
	
	
	\begin{figure}\centering
		\includegraphics[width=0.45\textwidth]{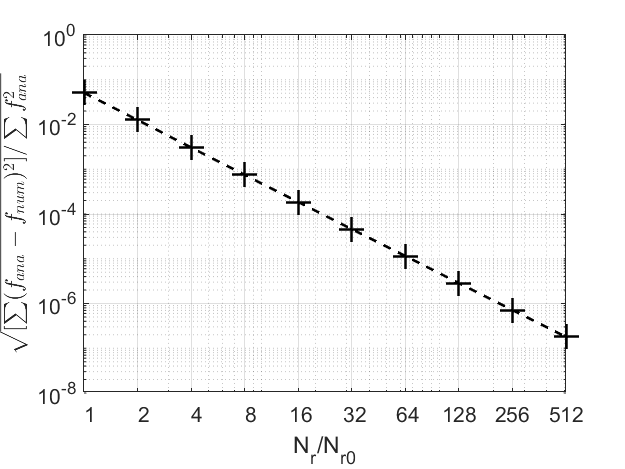}
		\caption{The relative error in the solution to the Amp\'ere's law versus different radial grid numbers using the Method of Manufactured Solutions, i.e., Eqs. \ref{eq:mms_rhs} and \ref{eq:mms_sol}, where $N_{r0}=10$.}
		\label{fig:mms_test}
	\end{figure}
	
	\begin{figure}\centering
		\includegraphics[width=0.7\textwidth]{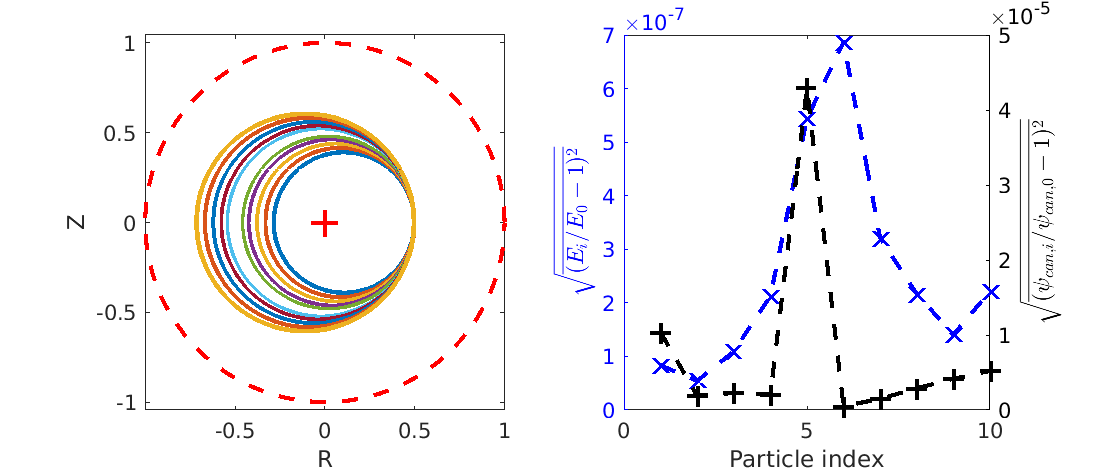}
		\includegraphics[width=0.7\textwidth]{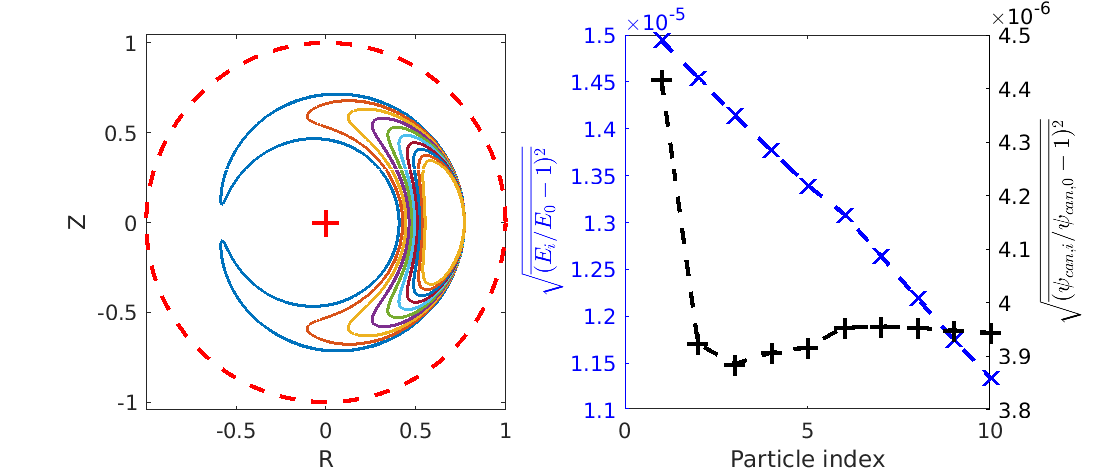}
		\caption{The guiding center trajectories and relative error of energy and toroidal canonical momentum for passing particles \revcolorb{(the upper row)} and trapped particles \revcolorb{(the lower row)}. 
		\revcolor{The root-mean-square (RMS) relative error is $\sqrt{\overline{(y_i/y_0-1)^2}}$, where $y_i$ is the signal at Step i, $y_0$ is the initial value and $\overline{(\ldots)}$ is the average over all steps. 
		The RMS relative error of $E$ and $P_\phi$ is smaller than $5\cdot10^{-5}$. 
		}
	}
		\label{fig:particle_EP_error}
	\end{figure}
	
	\begin{figure}\centering
		\includegraphics[width=0.48\textwidth]{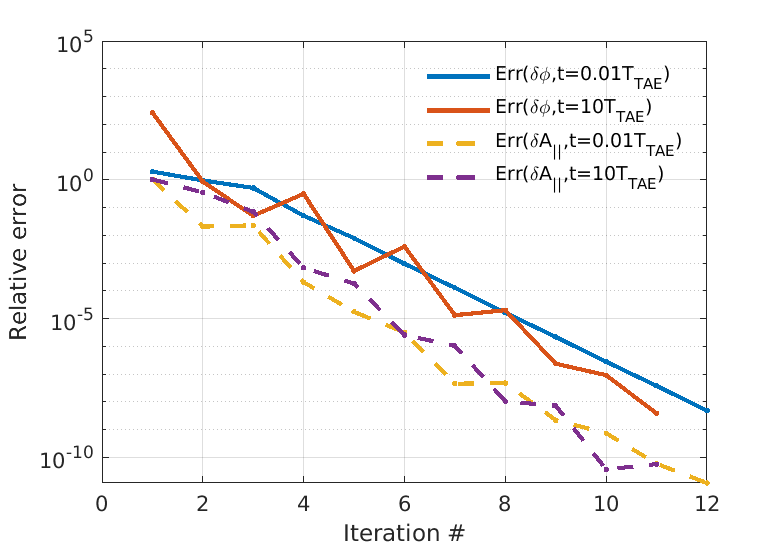}
		\caption{The convergence of $\delta\phi$ (lines) and $\delta{A}_\parallel$ (dashed lines) at the beginning ($t=0.01T_{TAE}$) and the end ($t=10T_{TAE}$) of EP driven TAE case in Section \ref{subsec:EP_TAE}.}
		\label{fig:convergence}
	\end{figure}
	
	\begin{figure*}\centering
		\includegraphics[width=0.325\textwidth]{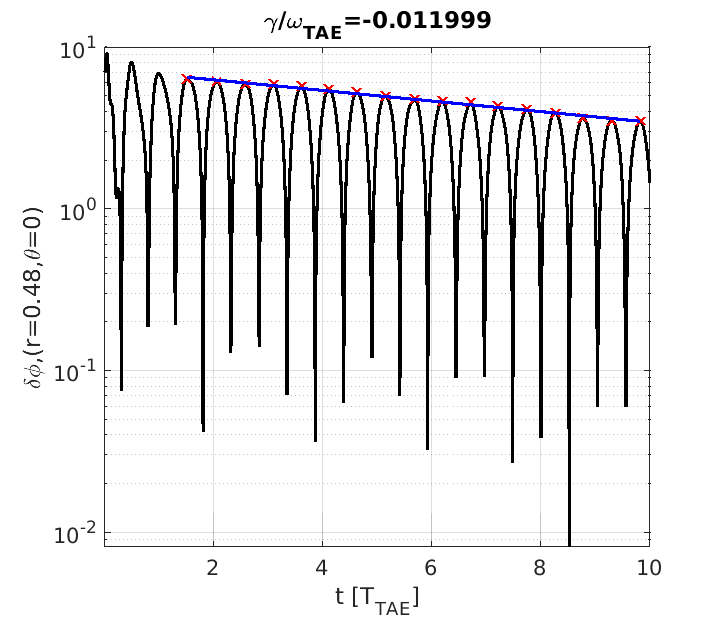}
		\includegraphics[width=0.65\textwidth]{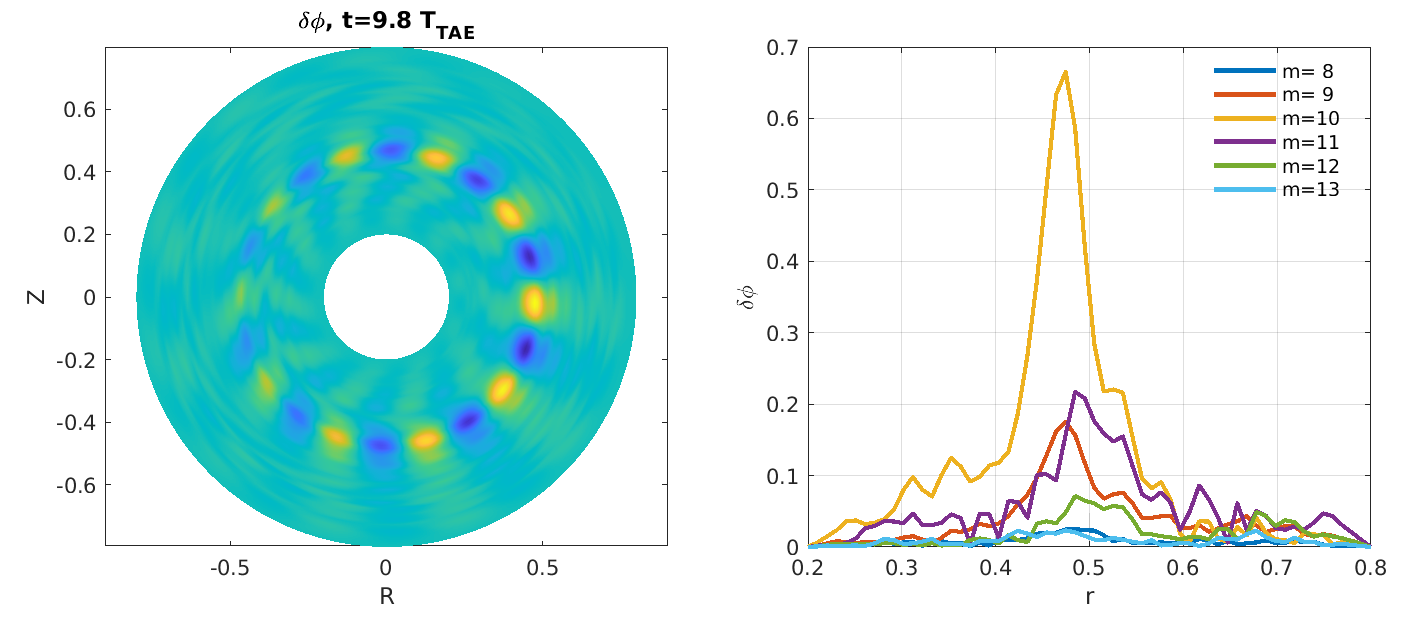}
		
		\includegraphics[width=0.33\textwidth]{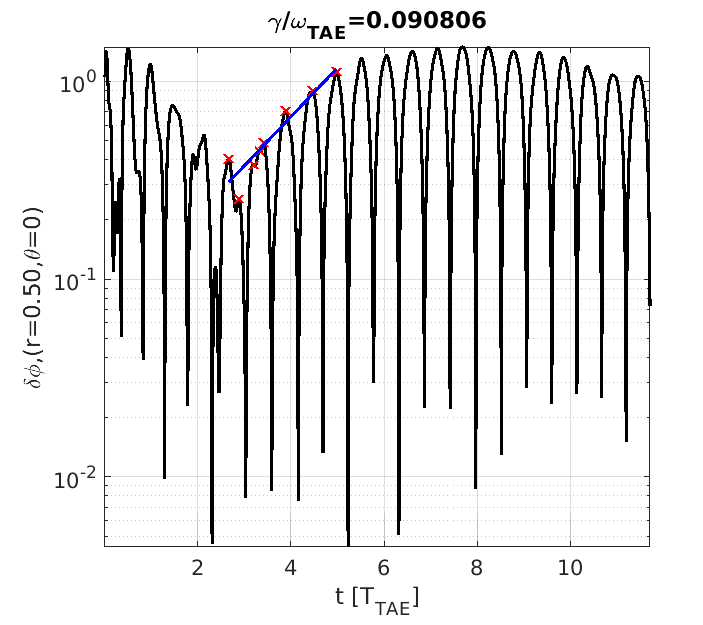}
		\includegraphics[width=0.65\textwidth]{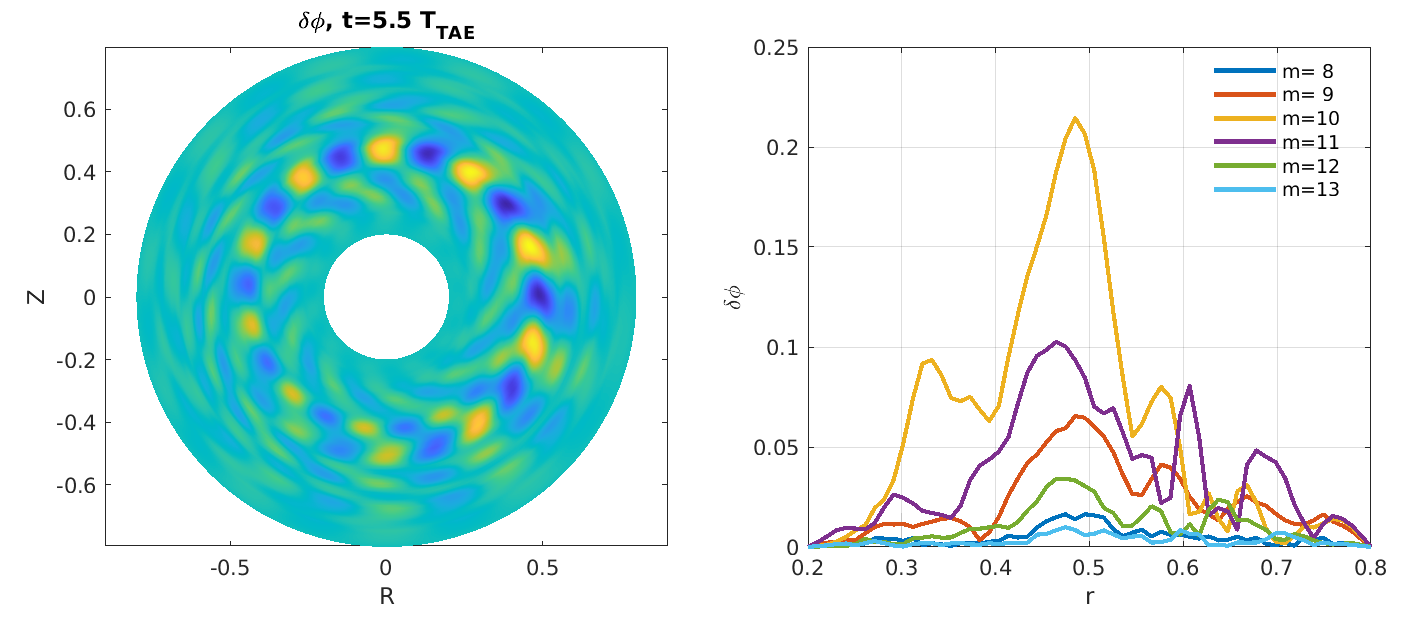}
		
		\caption{The first row: TAE damping w/o EPs; the second row: TAE driven by EPs. Left: the time evolution of the TAE. The blue line indicates the linear fit using the logarithmic amplitude peak values along $t$ during the selected time period (red crosses).  $\delta\phi(r_c,\theta=0)$ is normalized using $T_{e0}/e$ as adopted by other work \cite{biancalani2017nonlinear}. The 2D mode structure of the TAE $Re[\delta\phi]$ (middle column) and the radial structure of the different poloidal harmonics (right). The toroidal mode number $n=-6$, the electron to ion mass ratio $m_e/m_i=1/200$ for both cases. }
		\label{fig:tae_t1d_mode2d}
	\end{figure*}
	
	\ifthenelse{\equal{1}{1}}{}{
		
		\begin{figure}\centering
			\includegraphics[width=0.5\textwidth]{phic1d2.png}
			\caption{The time evolution of the TAE. The blue line indicates the linear fit using the logarithmic amplitude peak values along $t$ (red crosses).  $\delta\phi(r_c,\theta=0)$ is normalized using $T_{e0}/e$ as adopted by other work \cite{biancalani2017nonlinear}. }
			\label{fig:tae_damping}
		\end{figure}
		
		\begin{figure*}\centering
			\includegraphics[width=0.65\textwidth]{mode2d1.png}
			\caption{The 2D mode structure of the TAE $Re[\delta\phi]$ (left) and the radial structure of the different poloidal harmonics (right). }
			\label{fig:tae_damping2d}
		\end{figure*}

		\begin{figure}\centering
			\includegraphics[width=0.5\textwidth]{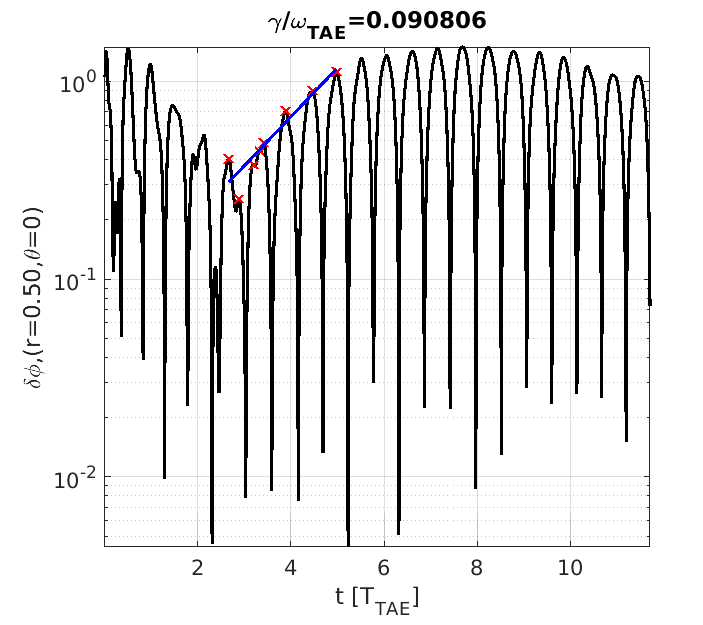}
			\caption{Time evolution of the EP driven TAE. The blue line indicates the linear fit during the growing phase ($t/T_{TAE}\in[2.5,5.5]$), using the logarithmic amplitude peak values along $t$ (red crosses).  $\delta\phi(r_c,\theta=0)$ is normalized using $T_{e0}/e$ as adopted by other work \cite{biancalani2017nonlinear}.}
			\label{fig:ep_tae}
		\end{figure}
		
		\begin{figure*}\centering
			\includegraphics[width=0.65\textwidth]{mode2d1ep.png}
			\caption{2D mode structures of the EP driven TAE mode$Re[\delta\phi]$ (left) and the radial structure of the different poloidal harmonics (right).   }
			\label{fig:ep_tae2d}
		\end{figure*}
		
	}
	
	\section{Summary and outlook }
	\label{sec:summery}
	
	In this work, an implicit full $f$  scheme has been developed for the electromagnetic particle simulations of the damping and the excitation of Alfv\'en modes. This work provides a potential method for EP transport simulations which is able to maintain the kinetic effects of all particles and the electromagnetic effect. 
	The main techniques have been developed as follows.
	\begin{enumerate}
		\item An analytical treatment has been derived for obtaining the implicit solution of the field-particle system, by linearizing the nonlinear implicit particle-field system, which gives a practical way to solve the nonlinear system, as shown in    Eqs. \ref{eq:dldtCN}, \ref{eq:dvdtCN},  \ref{eq:poissonCN}, \ref{eq:ampereCN}, \ref{eq:corrM} and \ref{eq:Mc_theory}.
		\item The mixed implicit-explicit scheme is developed to simulate the TAE by implicitly treating the \revcolor{fast scale} parallel motion, \revcolor{especially the parallel acceleration due to the perturbed field}, which is usually the most challenging when the particle mass is small, but treating the other parts explicitly.
	\end{enumerate}  
	The implicit scheme in this work shows the following performance in the study of Alfv\'en waves and EP physics.
	\begin{enumerate}
		\item Using the analytical derivation based implicit scheme, good convergence of the field-particle solver is demonstrated (Fig. \ref{fig:mms_test}).
		\item By applying \revcolorb{this} to the 1D shear Alfv\'en wave problem, this implicit scheme shows good energy conservation and capabilities of calculating the frequency and damping rate properly in a broad range of $\beta/M_e$ values, including the small electron mass condition (Fig. \ref{fig:saw_frequency_damping}).
		\item The application of this method to the TAE problem shows its applicability for electromagnetic simulations with/without EPs (Fig. \ref{fig:tae_t1d_mode2d}). The TAE mode structure distortion due to the non-perturbative effects of the EPs is observed, consistent with previous simulations \cite{biancalani2017nonlinear,wang2013radial} and theoretical studies \cite{ma2015global,lu2018kinetic,lu2019theoretical}.
	\end{enumerate}
	More dedicated studies related to the numerical performance of this implicit full $f$  scheme for the electromagnetic physics, \revcolor{such as the study of different discretization schemes for  more rigorous conservation properties,} will be addressed in future and physics problems such as the mode structure symmetry breaking and EP transport will be studied. The application of this method to the \revcolorb{whole plasma volume} using unstructured meshes \cite{lu2019development} or structured Bezier basis functions \cite{huysmans2007mhd}, is expected to enable more comprehensive studies of the global electromagnetic kinetic effects and edge physics. 
	
	\section*{Acknowledgments}
	Simulations in this work were performed on Max Planck Computing \& Data Facility (MPCDF). Discussions with and inputs from G. Huysmans, B. Sturdenvant, K. Kormann, A. Mishchenko, A. Bottino, F. Zonca, ORB5 team, EUTERPE team and HMGC team are appreciated by ZL. This work is supported by the EUROfusion Enabling Research Projects WP19-ER/ENEA-05 and WP19-ER/MPG-03. This work has been carried out within the framework of the EUROfusion Consortium and has received funding from the Euratom research and training programme 2014-2018 and 2019-2020 under grant agreement No 633053. The views and opinions expressed herein do not necessarily reflect those of the European Commission. \\
	
	
	\appendix
	\section{Field and guiding center equations in \revcolorb{$(r,\phi,\theta)$} coordinates}
	\label{appendix:gcmotion}
	In \revcolorb{$(r,\phi,\theta)$}, the Amp\'ere's law is written as 
	\begin{eqnarray}
	\label{eq:ampere_in_rtp}
	\left(L_{rr}-\frac{m^2}{r^2}\right)\delta{A}_{\parallel,m} &=&  C_A \delta J_{m}\;\;,\\
	L_{rr}&\equiv&\frac{\partial^2}{\partial r^2}+\frac{1}{r}\frac{\partial}{\partial r}\;\;,
	\end{eqnarray}
	where the perpendicular Laplacian operator has been approximated using that in $(r,\theta)$ plane, since $B_\theta/B_\phi=r/(qR)\ll1$. For the Poisson equation, the toroidal coupling is calculated using
	\begin{eqnarray}
	g_s&=& g_{s0}\frac{B_0^2}{B^2}\approx g_{s0}\left[1+2\epsilon_c\cos\theta\right]\;\;,
	\end{eqnarray}
	where $\epsilon_c=r/R_0$. 
	The Poisson equation is expressed as
	\begin{eqnarray}
	\label{eq:poisson_in_rtp}
	&&\left(L_{rr}-\frac{m^2}{r^2}\right)\delta{\phi}_{m} 
	+ \epsilon_c g_0 \left[ L_{rr}-\frac{m(m+1)}{r^2} \right]\delta\phi_{m+1} \nonumber\\
	&& + \epsilon_c g_0 \left[ L_{rr}-\frac{m(m-1)}{r^2} \right]\delta\phi_{m-1}
	= C_P \delta N_{m}\;\;,
	\end{eqnarray}
	where $g_0=\sum_s g_{s0}$.
	
	For guiding center's equations of motion, in $(r,\phi,\theta)$ coordinates, we have
	\begin{eqnarray}
	\frac{{d\bar{r}}_d}{dt} &=& \frac{M_sB_0\bar{\rho}_N}{\bar{e}_sB^3R} \frac{F}{r}\partial_\theta B \;\;, \\
	\frac{d\phi_d}{dt}      &=& \frac{M_sB_0\bar{\rho}_N}{\bar{e}_sB^3R} \frac{\partial_r\psi}{R}\partial_rB \;\;, \\
	\frac{d\theta_d}{dt}   &=& \frac{M_sB_0\bar{\rho}_N}{\bar{e}_sB^3R} \frac{F}{r}\partial_rB \;\;, 
	\end{eqnarray}
	\begin{eqnarray}
	\frac{d\delta{\bar{r}}}{dt} &=& \frac{B_0}{B}\bar{\rho}_N \left(  \frac{b_\phi}{r} \partial_\theta\delta G -\frac{b_\theta}{R} \partial_\phi\delta G  \right)\;\;, \\
	\frac{d\delta\phi}{dt}   &=&   \frac{B_0}{B}\bar{\rho}_N  \frac{b_\theta}{R} \partial_r\delta G\;\;, \\
	\frac{d\delta\theta}{dt} &=& - \frac{B_0}{B}\bar{\rho}_N  \frac{b_\phi}{r}   \partial_r\delta G\;\;, 
	\end{eqnarray}
	\begin{eqnarray}
	\dot{\bar{v}}_{\parallel0} &=& -\frac{\bar{\mu}\partial_r\psi}{R^2}\sin\theta\;\;,\\
	\nonumber\\
	\delta\dot{\bar{v}}_{\parallel} &=& -\frac{\bar{e}_s}{M_s}\left(\partial_\parallel\delta\bar{\phi}+\partial_t\delta \bar{A}\right) \;\;,
	\end{eqnarray}
	where $\delta G=\delta\bar{\phi}-\bar{v}_\parallel\delta A_\parallel$.
	
	
	%
	
\end{document}